\documentclass{aa}
\usepackage[dvips]{graphicx}
\usepackage{txfonts}
\usepackage{natbib}
\usepackage{color}
\bibpunct{(}{)}{;}{a}{}{,}    %to follow the A&A style 
\newcommand{\beq}{\begin{equation}}
\newcommand{\eeq}{\end{equation}}
\newcommand{\bea}{\begin{eqnarray}}
\newcommand{\eea}{\end{eqnarray}}
\newcommand{\url}[1]{{\tt #1}}

\def\gapp{\lower 3pt\hbox{${\buildrel > \over \sim}$}\ }
\def\lapp{\lower 3pt\hbox{${\buildrel < \over \sim}$}\ }

\newlength{\linwx}
\begin{document}
\title{Range of outward migration and influence of the disc's mass on the migration of giant planet cores}
%  \subtitle{empty}
%%
\author{
Bertram Bitsch \inst{1}
and
Wilhelm Kley  \inst{1}
}
\offprints{B. Bitsch,\\ \email{bertram.bitsch@uni-tuebingen.de}}
\institute{
     Institut f\"ur Astronomie \& Astrophysik, 
     Universit\"at T\"ubingen,
     Auf der Morgenstelle 10, D-72076 T\"ubingen, Germany
}
%%
%\date{February 25, 2009, revised May 27, 2009}
%%
\abstract
%%  Context
{The migration of planets plays an important role in the early planet-formation process. An important problem has been that standard migration theories predict very rapid inward migration, which poses problems for population synthesis models.
However, it has been shown recently that low-mass planets (20-30 $M_{Earth}$) that are still embedded in the protoplanetary disc can migrate outwards under certain conditions. Simulations have been performed mostly for planets at given radii for a particular disc model.
}
%%  Aims
{Here, we plan to extend previous work and consider different masses of the disc to quantify the influence of the physical disc 
conditions on planetary migration. The migration behaviour of the planets will be analysed for a variety of positions in the disc.
}
%% Methods
{We perform three-dimensional (3D) radiation hydrodynamical simulations of embedded planets in protoplanetary discs.
We use the explicit-implicit 3D hydrodynamical code {{\tt NIRVANA}} that includes full tensor viscosity, and
implicit radiation transport. For planets on circular orbits at various locations we measure the radial dependence of the torques
for three different planetary masses.
}
%% Results
{For all considered planet masses (20-30 $M_{Earth}$) in this study we find outward migration within a limited radial range of the disc, typically from about 0.5 up to 1.5-2.5 $a_{Jup}$. Inside and outside this interval, migration is inward and given by the Lindblad value for large radii. Interestingly, the fastest outward migration occurs at a radius of about $a_{Jup}$ for different disc and planet masses. Because outward migration stops at a certain location in the disc, there exists a zero-torque distance for planetary embryos, where they can continue to grow without moving too fast. For higher disc masses ($M_{disc} > 0.02 M_\odot$) convection ensues, which changes the structure of the disc and therefore the torque on the planet as well.
}
%% Conclusions
{
Outward migration stops at different points in the disc for different planetary masses, resulting in a quite extended region where the formation of larger cores might be easier. In higher mass discs, convection changes the disc's structure resulting in fluctuations in the surface density, which influence the torque acting on the planet, and therefore its migration rate. Because convection is a 3D effect, 2D simulations of massive discs with embedded planets should be handled with care.
}
\maketitle
\markboth
{Bitsch \& Kley: Range of outward migration and influence of the disc's mass on planetary migration}
{Bitsch \& Kley: Range of outward migration and influence of the disc's mass on planetary migration}

\section{Introduction}
\label{sec:introduction}
Planetary migration of embedded low-mass planets in isothermal discs indicates inward migration, so that the planet might be lost in the star before the accretion disc is gone \citep{2002ApJ...565.1257T}. Recent studies (starting with the work of \citet{2006A&A...459L..17P}) have shown that the inclusion of radiation transport in planet-disc interaction studies resulted in a slowed down or even reversed migration \citep{2008ApJ...672.1054B, 2008A&A...485..877P, 2008A&A...478..245P, 2008A&A...487L...9K, 2009A&A...506..971K, 2010MNRAS.408..876A, 2011arXiv1103.3502A}.

All authors agree that the inclusion of radiation transport is an important effect that should be considered, however, not all authors find outward migration. The efficiency of outward migration  depends on the magnitude of positive corotation torques. These are determined by the local entropy and vortensity gradients. For isothermal and adiabatic discs these gradients cannot be maintained and so-called torque saturation reduces the corotation effects. However, the inclusion of radiative transport and viscosity prevents saturation, and the negative Lindblad torques (caused by the spiral arms) can be overwhelmed by the positive contributions from the corotation region.   The magnitude of the effect depends on the planetary mass \citep{2008A&A...487L...9K, 2009A&A...506..971K} and the temperature gradient in the disc. For example, \citet{2011arXiv1103.3502A} found for temperature slopes of $\beta > 1.0$ (with $T \propto r^{-\beta}$) that outward migration is possible even for p
 lanets with up to $50 M_{Earth}$ in 3D simulations. The migration rate increases with an increasing temperature slope. These results are also reflected by the theoretical analysis from \citet{2010ApJ...723.1393M}. Recently, improved theoretical torque formulae for low-mass  planets embedded in an adiabatic disc have been presented by \citet{2009ApJ...703..857M} and \citet{2010MNRAS.401.1950P}.
The most recent improvements consider the necessary inclusions of radiative diffusion and viscosity
\citep{2010ApJ...723.1393M, 2011MNRAS.410..293P}. These have been developed for 2D discs where
the diffusive effects can operate only in the disc's plane. Interestingly, radiative cooling alone can lead to similar, even stronger effects \citep{2008A&A...487L...9K}. 
 
A very important open question is how far the planet will move outwards in a fully radiative disc. When the protoplanets are stopped from migrating outwards at a certain point in the disc, a protoplanet moving inwards from farther out will stop at the same location in the disc. Therefore, a planetary trap inside the disc can be created, which can act as an area of planetary mergers, leading to larger cores. A planetary trap inside the disc created by surface density changes can function as a feeding zone to planetary cores \citep{2008A&A...478..929M}, but it is unclear how realistic these surface density changes are. The inclusion of radiation transport/cooling in a disc might provide such a trap in a normal disc structure, at least for planets within a given mass range.

In \citet{2011ApJ...728L...9S} the authors show in N-body simulations that because of the inclusion of unsaturated type-I-migration \citep{2010MNRAS.401.1950P} large planetary cores (up to $10 M_{Earth}$) can form in protoplanetary discs well before the disc is accreted onto the star.

In population synthesis models that include the standard migration prescription for isothermal discs the migration is too fast, such that the outcome distribution does not match the observed one.
Specifically, the type-I-migration rate should be $10$ to $1000$ times slower than expected from linear analysis \citep{2004A&A...417L..25A, 2008ApJ...673..487I, 2009A&A...501.1139M}. As pointed out above, this could be provided by the inclusion of radiation transport in hydrodynamical type-I-migration simulations. An updated version of the type-I-migration, by using the formula from \citet{2010MNRAS.401.1950P}, in population synthesis models shows very promising results \citep{2011arXiv1101.3238M}, but additional simulations are needed to verify these assumptions.

In our previous simulations and studies \citep{2009A&A...506..971K, 2010A&A.523...A30, 2011A&ABitschKley} we have only considered one standard disc model with a given mass. In reality, of course, protoplanetary discs can have a variety of masses. A fully radiative disc will settle in an equilibrium state between viscous heating and radiative transport/cooling. Given that all other physical parameters are fixed, the resulting disc structure only depends on the disc's mass and viscosity, because a more massive or more viscous disc is able to heat the disc more effectively. A more massive disc therefore influences the migration rate of embedded planets. Here we focus on low-mass planets that can undergo outward migration in fully radiative discs.

In this paper we extend previous studies and investigate the possible radial range over which outward migration can occur and analyse the influence of the disc's mass on the migration in detail. An important effect is the onset of convection in the disc, which becomes stronger for more massive discs.
 
In Section \ref{sec:setup} we give a short overview of our code and numerical methods, where more details can be found in \citep{2009A&A...506..971K}. The influences of the distance of the embedded planet to the central star is discussed in Section \ref{sec:outwardrange}. We then analyse the influence of the disc's mass on the disc's structure (density, temperature, aspect ratio) and then the influence on migration of embedded low-mass planets in Section \ref{sec:discmass}. Convection in the disc is also briefly discussed in Section \ref{sec:discmass}. We then summarize and conclude in Section \ref{sec:Sumcon}.

\section{Setup of the simulations}
\label{sec:setup}

The protoplanetary disc is modelled as a three-dimensional (3D), non-self-gravitating gas whose motion is described by the Navier-Stokes equations. We treat the disc as a viscous medium, where the dissipative effects can then be described via the viscous stress-tensor approach. We also assume that the heating of the disc occurs solely through internal viscous dissipation and ignore the influence of additional energy sources (e.g. irradiation form the central star). This internally produced energy is then radiatively diffused through the disc and eventually emitted from its surface. For this process we use the flux-limited diffusion approximation, which allows us to treat the transition from optically thick to thin regions as an approximation. A more detailed description of the modelling and the numerical methodology is provided in our previous papers \citep{2009A&A...506..971K, 2010A&A.523...A30, 2011A&ABitschKley}, and for that purpose we limit ourselves here to present only
  the most necessary information.

\subsection{Physical setup}

We solve the Navier-Stokes equations numerically using a spatially second order finite volume method based on the code 
{\tt NIRVANA} \citep{1997ZiegYork}, with implicit radiative transport in the flux-limited diffusion approximation and the {\tt FARGO} \citep{2000A&AS..141..165M} extension as described in \citet{2009A&A...506..971K}.
We use a spherical polar coordinate system ($r, \theta, \phi$), where the computational domain consists of a complete annulus 
($0 \leq \phi \leq 2 \pi$) of the protoplanetary disc centred on the star, extending from $r_{min}$ to $r_{max}$, which are determined by the location of the planet. 
For symmetry reasons and because we only use non-inclined planets, we restrict ourselves in the standard setup to the upper half of the disc. Hence, in the vertical direction the annulus extends from the equator up to $7^\circ$ above the disc's midplane, meaning $83^\circ < \theta < 90^\circ$. 
Here $\theta$ denotes the polar angle of our spherical polar coordinate system measured from the polar axis. For the study of convection
we use in addition a two-sided disc, see below.
The central star has one solar mass $M_\ast = M_\odot$, and the total disc mass inside [$r_{min}, r_{max}$] is $M_{disc} = 0.01 M_\odot$, unless stated otherwise in Section \ref{sec:discmass}. For our radiative models the temperature and subsequently $H/r$ is calculated self-consistently from the equilibrium structure given by the viscous internal heating and radiative diffusion. We note that for given physics (equation of state, opacity, viscosity) the structure of the disc is solely dependent on its mass, and this is one aspect that we will investigate in this paper.

For the radiative transport we use a one-temperature approach and apply the flux-limited diffusion approximation using analytic Rosseland opacities, for details see \citet{2009A&A...506..971K}.  To close the basic system of equations we use an ideal gas equation of state with constant mean molecular weight $\mu=2.35$ for
a standard solar mixture. The adiabatic index is $\gamma =1.4$. 
For the present study, we use a constant kinematic viscosity coefficient with a value of $\nu = 10^{15}$\,cm$^2$/s, a value that relates to an equivalent $\alpha = 0.004$ at $5.2 AU$ for a disc aspect ratio of $H/r = 0.05$, where $\nu = \alpha H^2 \Omega_K$. 
In standard dimensionless units with $r_0=a_{Jup}=5.2 AU$ and $t_0 = \Omega_K(r_0)^{-1}$ we have $\nu = 10^{-5}$. 

\subsection{Calculation setup}

Our coordinate system rotates at the initial orbital frequency of the planet (at $r=r_P$). We use an equidistant spherical grid in $r,\theta,\phi$ with a resolution of ($N_r,N_\theta,N_\phi)=(266,32,768)$ active cells for our simulations. At $r_{min}$ and $r_{max}$ we use damping boundary conditions for all three velocity components to minimize disturbances (wave reflections) at these boundaries. The velocities are relaxed towards their initial state on a time scale of approximately the local orbital period. The angular velocity is relaxed towards the Keplerian values, while the radial velocities at the inner and outer boundaries vanish. Reflecting boundary conditions are applied for the density and temperature in the radial directions. We apply periodic boundary conditions for all variables in the azimuthal direction. In the vertical direction we set outflow boundary conditions for $\theta_{min} = 83^\circ$ (the surface of the computational domain). 

In our previous work, we have discussed the calculation of the torque acting on a planet in great detail. Outward migration seems only possible and is strongest when the planet is on nearly circular orbits in the midplane of the disc \citep{2010A&A.523...A30, 2011A&ABitschKley}. Additionally, we stated in \citet{2011A&ABitschKley} that the measured migration rate from planets on fixed orbits is equivalent to the migration rate determined from moving planets in discs. 
Hence, we consider here primarily planets on fixed circular orbits in the midplane of the disc, and calculate the torque acting on the planet, because the torque represents a direct measurement of migration in this case. 
For a one-sided disc we use symmetric boundaries at $\theta_{max} = 90^\circ$ (the disc's midplane). To correctly see the influence of convection in the disc we use two-sided discs for some high-mass discs. For these simulations we used outflow boundaries for both $\theta_{min}$ and $\theta_{max}$.

The models are initialized with constant temperatures on cylinders with a profile $T(s) \propto s^{-1}$ with $s=r \sin \theta$.
The initial vertical density stratification is approximately given by a Gaussian where the radial surface density follows a $\Sigma(r) \propto \, r^{-1/2}$ profile. In the radial and $\theta$-direction we set the initial velocities to zero, while for the azimuthal component the initial velocity $u_\phi$ is given by the equilibrium of gravity, centrifugal acceleration, and the radial pressure gradient. This corresponds to the equilibrium configuration for a purely isothermal disc. Before embedding the planet, we bring the disc into radiative equilibrium by performing first 2D axisymmetric simulations in the $r-\theta$ plane. This takes about 100 orbits. We then extend this model to a full 3D simulation by expanding the grid into the $\phi$-direction, and the planet is embedded.

For the gravitational potential of the planet we utilize the cubic potential, where the potential is exact beyond a transition (smoothing) radius $r_{sm}$ and smoothed by a cubic polynomial inside \citep{2006A&A...445..747K,2009A&A...506..971K}. Here we use  $r_{sm}= 0.5 R_{H}$, where $R_H$ is the Hill radius of the planet.

The torques acting on $20$, $25$, and $30 M_{Earth}$ planets are calculated to determine the direction of migration. The gravitational torques acting on the planet are calculated by integrating over the whole disc, where we apply a tapering function to exclude the inner parts of the Hill sphere of the planet \citep{2008A&A...483..325C}. This torque-cutoff is necessary to avoid strong, probably noisy contributions from the inner parts of the Roche lobe and to disregard material that is gravitationally bound to the planet \citep{2009A&A...502..679C}. Here we assume (as in our previous papers) a transition radius of $0.8$ Hill radii.

\section{Range of outward migration}
\label{sec:outwardrange}

Previous simulations by several authors \citep{2008ApJ...672.1054B, 2008A&A...485..877P, 2008A&A...478..245P, 2008A&A...487L...9K, 2009A&A...506..971K} indicated that outward migration of low-mass planets may be possible during an early evolutionary state of planet formation. However, because the simulations dealt mostly with planets at a given distance from the star, typically 5.2 AU, the radial range over which the migration may be directed outwards has not been addressed in great detail. In \citet{2010A&A.523...A30} we obtained some results for moving planets but the extent of the outward migration remained unclear.

In order to address this problem, we simulate $20, 25$, and $30 M_{Earth}$ planets on fixed circular orbits embedded in fully radiative discs at various distances from the host star. The planet's semi-major axis $r_p$ lies in a range of $0.6 \leq r_P \leq 5.0 r_0$, where $r_0 = a_{Jup}=5.2 AU$. With increasing distance from the star, the density and temperature of the disc decrease and at some point in the disc the conditions for outward migration might not be fulfilled anymore. 

For this set of simulations with different planet locations we use a disc setup with a density profile such that the total disc mass equals $M_{Disc}=0.01 M_\odot$ for a planet at $r_p = 1$ and $r_{min} = 0.4$ and $r_{max} = 2.5$ in units of $r_0$.
The planets are embedded in a way that the distance to the inner edge is always the planets location divided by $2.5$, while the distance to the outer edge is the planet's location to the star multiplied by $2.5$. This setup ensures that the radial boundaries are always far enough away so they do not influence our results of embedded planets. Since the overall surface density profile ($\Sigma \propto r^{-0.5}$) of the different disc models refers to the same physical situation, the total disc mass in the computational domain changes in the same way as the computed domain changes its size. The surface density at a given distance to the central star is constant for all disc models. The rotation frequency of the grid matches the rotation speed of the planet, so that the planet remains at a fixed position inside the computational grid at all times.

\begin{figure}
 \centering
 \includegraphics[width=0.9\linwx]{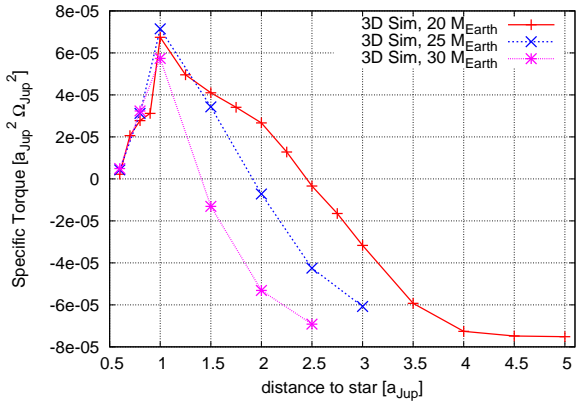}
 \caption{Torque acting on planets with three different masses embedded in fully radiative discs located at various distances from the star 
 (red for $20 M_{Earth}$, blue for $25 M_{Earth}$ and purple for $30 M_{Earth}$). The torques are plotted when planet and disc have reached a quasi-stationary equilibrium where the torque acting on the planet is approximately constant in time.
   \label{fig:MigTorqueMass}
   }
\end{figure}

In Fig.~\ref{fig:MigTorqueMass} the specific torque (per planet mass) acting on the three planetary masses at different distance from the central star is displayed. For all planet masses the most extended positive torque (indicating outward migration) is around $r \approx 1.0 a_{Jup}$. At longer distances to the central star, the torque acting on the planets decreases continuously to negative torques, and this transition from positive to negative torques occurs at larger distances for lower planet masses. For the lowest mass planet with $20 M_{Earth}$ the transition is at $r \approx 2.4 a_{Jup}$ (zero-torque distance to the central star). With even longer distances the torques remain negative but with diminishing strength, indicating inward migration. For higher planetary masses ($25$ and $30 M_{Earth}$) the zero-torque distance is decreasing ($1.9$ and $1.4 a_{Jup}$). For shorter distances to the central star, inside the maximum, the torque acting on the planet is reduced again until it reaches about zero for $r_P = 0.5 a_{Jup}$ for all planetary masses.

\begin{figure}
 \centering
 \includegraphics[width=0.9\linwx]{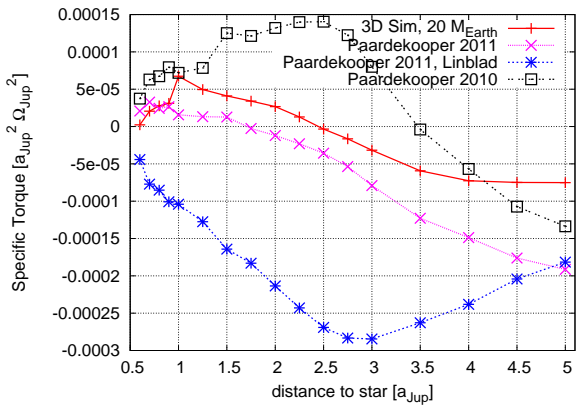}
 \caption{Torque acting on a $20 M_{Earth}$ planet in a fully radiative disc as a function of distance from the star (red plus signs).
  Additional curves indicate the results from the torque formula published in \citet{2010MNRAS.401.1950P} (black squares) 
   and \citet{2011MNRAS.410..293P} (purple crosses). 
  Additionally we also plot the linear Lindblad torque by the \citet{2011MNRAS.410..293P} formula (blue asterisks).
   \label{fig:MigTorquePaar}
   }
\end{figure}

In Fig.~\ref{fig:MigTorquePaar} the torque for the $20 M_{Earth}$ planet is again displayed together with semi-analytical results from \citet{2010MNRAS.401.1950P, 2011MNRAS.410..293P}. 
The black boxes refer to the torque formula presented in \citet{2010MNRAS.401.1950P}, which applies for the unsaturated torques in adiabatic discs. It is given by
\begin{equation}
\label{eq:paar09}
     \gamma \Gamma_{tot} / \Gamma_0 = -2.5 - 1.7 \beta + 0.1 \alpha + 1.1 (1.5-\alpha) + 7.9 \xi / \gamma \ ,
\end{equation}
with $\alpha$ and $\beta$ being the slope of the radial surface density and midplane temperature profile, respectively
\begin{equation}
         \Sigma(r)  \propto r^{-\alpha}, \quad   T(r) \propto r^{-\beta}.
\end{equation}
To calculate the semi-analytic results in the plot we use a constant $\alpha = 0.5$, and $\beta$ changes from about 
$1.7$ at $r=1.0 a_{Jup}$ to $0.33$ at $r=5.0 a_{Jup}$, see also Fig.~\ref{fig:MassRTHR} below.
The slope of the entropy profile is then given by $\xi = \beta - (\gamma -1) \alpha = 1.5$ at $r=1.0 a_{Jup}$.
The torque is normalized to
\[
      \Gamma_0 = \left(\frac{q}{h}\right)^2 \Sigma_P r_p^4 \Omega_P^2 \ ,
\]
with $q$ the planet/star mass ratio, $h$ the relative disc height $(H/r)$, $\Sigma_P$ the surface density at the planet's location
and $\Omega_P$ the rotation frequency of the planet in the disc.
The vertical height $H$ of the disk is defined as $H = c_s/\Omega$ where the midplane values are used for the adiabatic sound speed $c_s$ and the
angular velocity $\Omega$. 

At $r=1.0 a_{Jup}$ the formula from \citet{2010MNRAS.401.1950P} agrees well with our 3D simulations. 
As can be inferred from Eq.~(\ref{eq:paar09}), the theoretical torque could never become negative for constant slopes ($\alpha, \beta, \xi$).
However, as $\beta$ varies with $r$, a change from positive to negative torques is possible and indeed occurs 
at $r \approx 3.5$ (black squares in Fig.~\ref{fig:MigTorquePaar}).
However, in the range of $1.2 < r < 3.8$ the torque formula predicts a much higher torque than our 3D simulations, as is the case
for larger distances to the central star, where the formula suggests a larger negative torque compared with our results. 
It should be noted that Eq.~(\ref{eq:paar09}) includes Lindblad and corotation torques where the latter include contributions from the vorticity as well as entropy gradient. One should also be aware that Eq.~(\ref{eq:paar09}) is valid only at the beginning of the evolution when the torques acting on the planet are not saturated. However, at later times the flow settles to an equilibrium state where the torques saturate and Eq.~(\ref{eq:paar09}) is not valid any more.

In Fig.~\ref{fig:MigTorquePaar} we additionally plot results from the improved formula in \citet{2011MNRAS.410..293P} (purple crosses).
Its derivation is quite complex and we give a brief summary in the appendix. The formula captures the behaviour of the torque caused by Lindblad resonances and horseshoe drag on low-mass planets embedded in gaseous discs in the presence of viscous and thermal diffusion \citep{2011MNRAS.410..293P}.
The new formula of \citet{2011MNRAS.410..293P} including diffusion provides a slightly better fit compared to the adiabatic \citet{2010MNRAS.401.1950P} formula. 
The torque is positive for $r < 1.8 a_{Jup}$ and becomes negative for larger distances to the central star. 
In general the formula captures the trend of our simulations, but lies consistently below our radiative disks. 
Also the zero-torque equilibrium radius differs.  At the often used reference radius $r=a_{Jup}$ the torque formula 
and our 3D simulations differ by a factor of about three.
For $r > 2.5 a_{Jup}$ our simulations show a negative torque acting on the planet, which continues to grow until about $r \approx 4.0 a_{Jup}$. 
For these larger radii the difference between our results and the analytic formula increases.

\begin{figure}
 \centering
 \includegraphics[width=0.9\linwx]{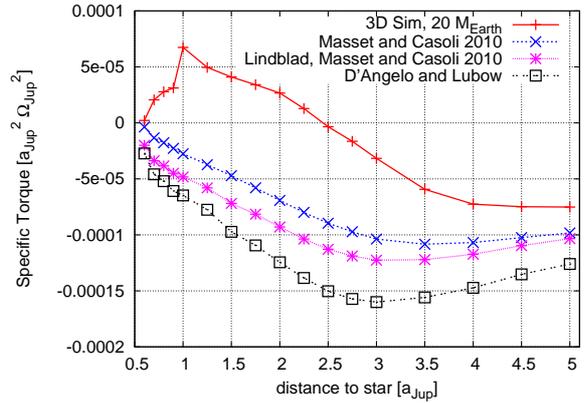}
 \caption{Torque acting on a $20 M_{Earth}$ planet in fully radiative discs as a function of distance from the star.
  The additional curves indicate the estimates from the theoretical formula for viscous, diffusive disks by \citet{2010ApJ...723.1393M} 
  and for isothermal disks by \citet{2010ApJ...724..730D}.
   \label{fig:MigTorqueMasset}
   }
\end{figure}

In Fig.~\ref{fig:MigTorquePaar} the Lindblad torque according to \citet{2011MNRAS.410..293P} is also displayed (blue). 
For larger distances to the star, the disc becomes thinner and the effects of the corotation torque become less important. 
The total torque should therefore converge towards the Lindblad torque, however the torques from our simulations differ by a factor 
of three at $r=5.0 a_{Jup}$ with the Lindblad torque of \citet{2011MNRAS.410..293P}. 
As the Lindblad torque is dependent on the temperature gradient $\beta$, Eq.~(\ref{eq:paar09}), and on a pre-factor,
a reduction of the pre-factor as in \citet{2010ApJ...723.1393M} reduces the Lindblad torque (see also Fig.~\ref{fig:MigTorqueMasset}).

In Fig.~\ref{fig:MigTorqueMasset} we also plot the results from \citet{2010ApJ...723.1393M}, who provided an alternative recipe for planetary migration in viscous discs with thermal diffusion. We do not cite the full formulae here, due to their complexity. In paragraph 8 of \citet{2010ApJ...723.1393M}, a summary of the torque formula is given. 
Additionally, the isothermal results of the 3D simulations of \citet{2010ApJ...724..730D} are plotted.

Compared to our 3D results the torque formula of \citet{2010ApJ...723.1393M} leads to qualitatively different results.
The torque predicted by their formula is negative for all investigated distances to the central star.
For large distances the torque of \citet{2010ApJ...723.1393M} matches better with our simulations, but is still about $35\%$ too low. 
The Lindblad torque of \citet{2010ApJ...723.1393M} is smaller compared to the Lindblad torque in \citet{2010MNRAS.401.1950P} 
due to a different pre-factor in front of the temperature slope $\beta$ (in Eq.~(\ref{eq:paar09})).
The 3D isothermal formula by \citet{2010ApJ...724..730D} shows a larger Lindblad torque compared to our simulations and to \citet{2010ApJ...723.1393M}.
But as the formula is for isothermal discs only, the torque has to be reduced by a factor of $\gamma$ compared to our radiative simulations 
or compared to the torque from \citet{2010ApJ...723.1393M}. Taking this into account the torques match quite well for large distances to the central star.

To test our implementation of the torque formula by \citet{2010ApJ...723.1393M}, we applied it to the results of  \citet{2011arXiv1103.3502A} using our
temperature gradient (see their Fig.~3).
However, in contrast to \citet{2011arXiv1103.3502A} we find negative torques when applying that formula.
Apparently, when comparing their numerical results to the anytical formula  by \citet{2010ApJ...723.1393M},
Ayliffe \& Bate applied the latter with a somewhat larger viscosity (approx. by a factor of $2\pi$) than quoted in text.
This resulted in the positive torque quoted in their paper. (B. Ayliffe, private communication).

There are many reasons for the differences between the simulations and the formulae. 
The formulae in \citet{2010MNRAS.401.1950P,2011MNRAS.410..293P} and in \citet{2010ApJ...723.1393M} were derived for 2D discs only,
but the horseshoe drag can be even stronger in three dimensions, as shown for isothermal discs in \citet{2006ApJ...652..730M}. 
The radiative diffusion is also most effective in the vertical direction, meaning that the two-dimensional approximation where diffusion is is restricted to the orbital plane, does not capture the true physical effects \citep[see also][]{2008A&A...487L...9K}.
The formulae were derived for background gradients of temperature and surface density, but as the disc with an embedded planet evolves, the structure of the disc changes and the basic assumptions (gradients in temperature, density, and so on) used to derive the formulae might not be valid any more. This is in particular true for planets in the mass range studied here, because the theory has been developed for lower mass planets, which do not alter their surroundings significantly. A more detailed discussion about the smoothing of the planet in \citet{2011MNRAS.410..293P} can be found in Appendix \ref{app:comp}. Nevertheless, there seems to be a very rough qualitative agreement between the improved torque formula of \citet{2011MNRAS.410..293P} and our numerical results.

\begin{figure}
 \centering
 \includegraphics[width=0.9\linwx]{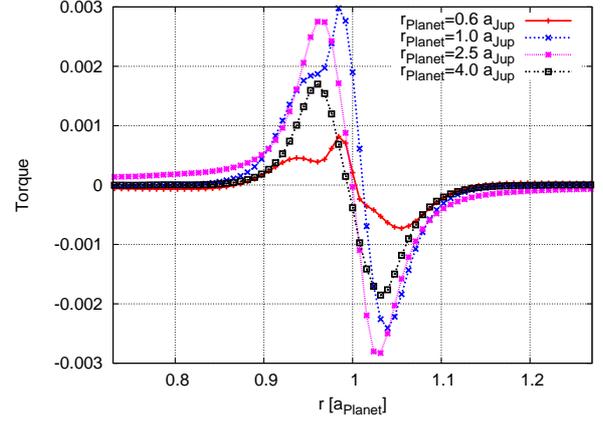}
 \caption{Radial torque density $\Gamma (r)$ acting on $20 M_{Earth}$ planets at $r_P=0.6$, $r_P=1.0$, $r_P=2.5$ and $r_P=4.0 a_{Jup}$. The distances have been scaled in $r$-direction to the actual planet location $r_P$.The snapshots of the torque density have been taken in the equilibrium state at $80$ Jupiter orbits.
   \label{fig:MigGamma3D}
   }
\end{figure}

In Fig.~\ref{fig:MigGamma3D} the radial torque density $\Gamma (r)$ is displayed for $20 M_{Earth}$ planets at $r_P=0.6$, $r_P=1.0$, $r_P=2.5$ and $r_P=4.0 a_{Jup}$, where the radius is normalized to the actual planetary distances to allow a direct comparison. In all curves the contributions by the Lindblad torques are clearly visible, positive for $r<1$ and negative for $r>1$. The contribution responsible for the torque reversal, the 'spike' just inside $r=1$, is visible only for the $r_P = 0.6$ and $r_P = 1.0$ locations, which both show an overall positive total torque. Even though the torque acting on the $r_P=0.6 a_{Jup}$ planet is much reduced, the spike in the torque distribution is clearly visible. In \citet{2009A&A...506..971K} we discussed the origin of the spike for the $r_P=1.0 a_{Jup}$ planet. It is an indicator for a density enhancement ahead of the planet just inside of $r_P$, and thus creating a positive torque. 

For the $r_P=2.5 a_{Jup}$ case the corotation spike in the fully radiative case in the torque density is not visible any more, only the Lindblad torques are visible. The resulting total torque is about zero, which indicates that the Lindblad torques are just counterbalanced by the corotation effects. For even longer distances to the central star, the torque is identical to the (negative) Lindblad torque, indicating inward migration (see also Fig.~\ref{fig:MigTorqueMass}).

\begin{figure}
 \centering
 \includegraphics[width=0.795\linwx]{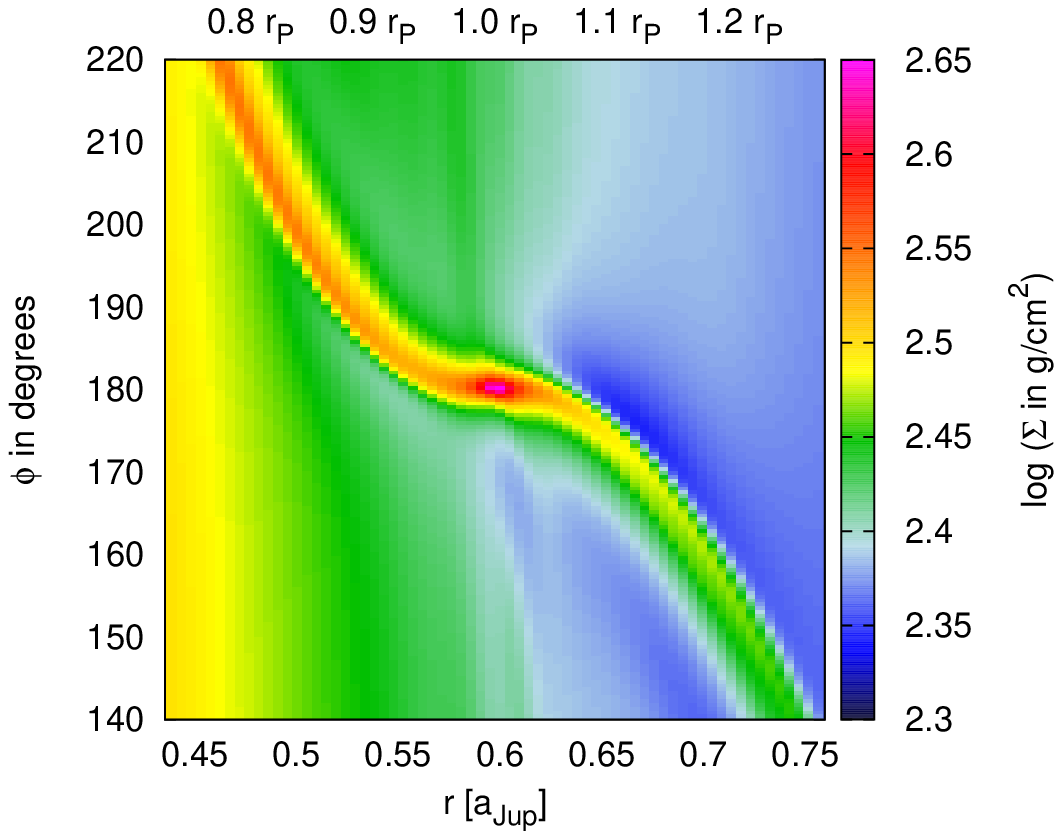}
 \includegraphics[width=0.795\linwx]{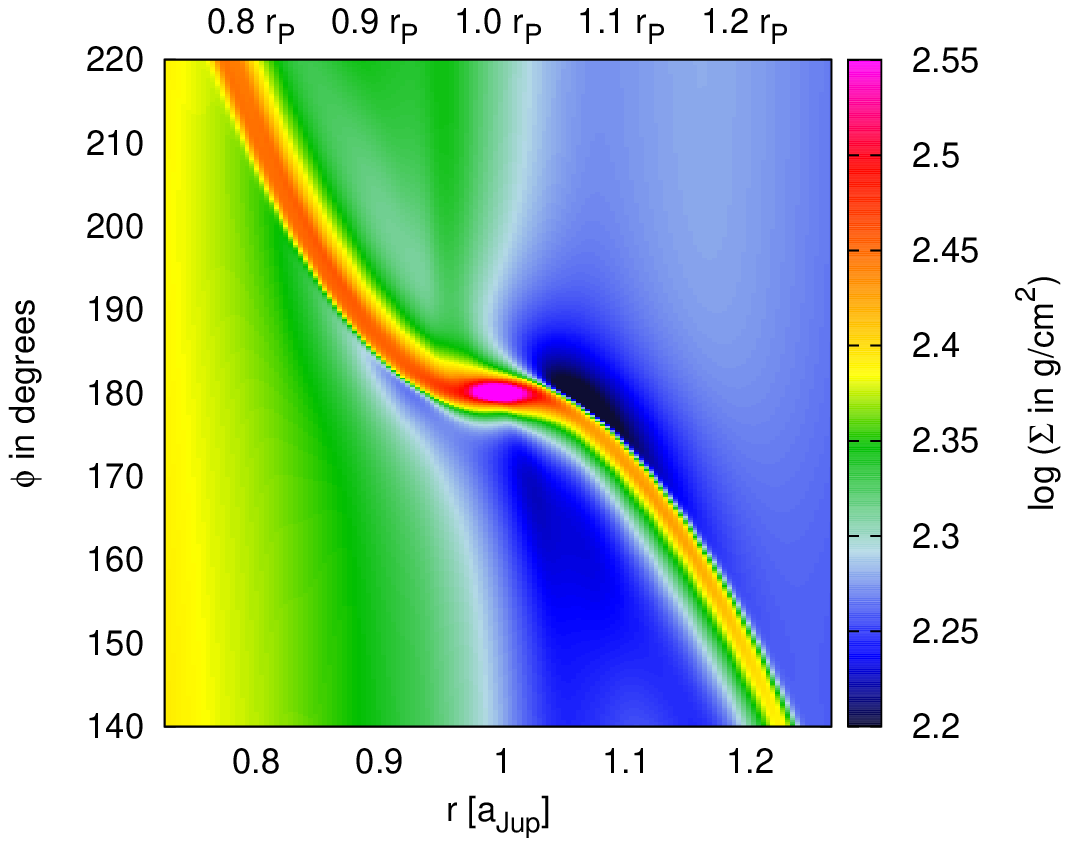}
 \includegraphics[width=0.795\linwx]{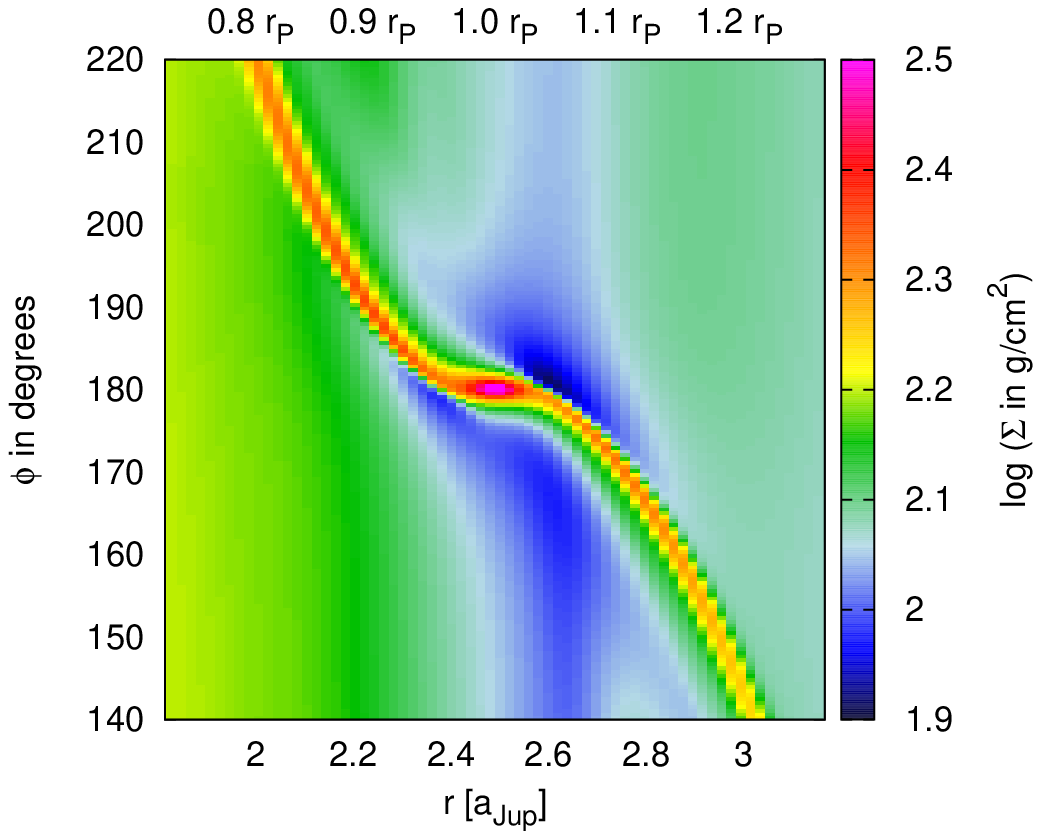}
 \includegraphics[width=0.795\linwx]{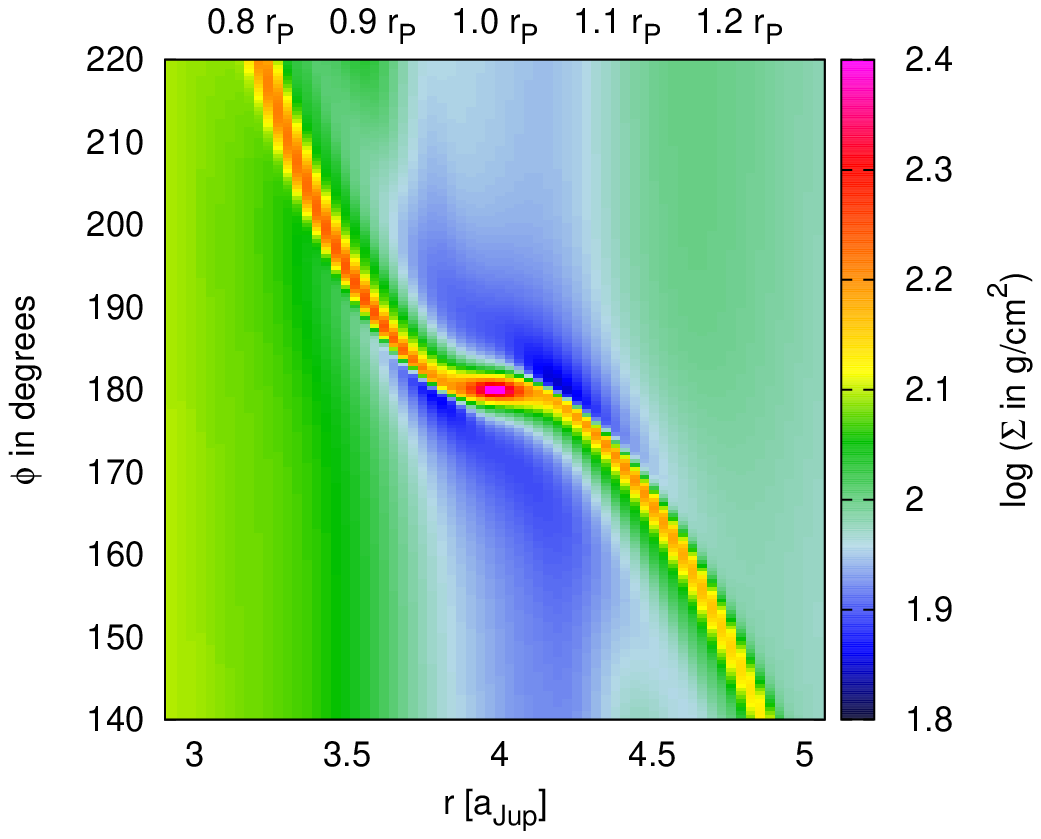}
 \caption{Surface density maps for a $20 M_{Earth}$ planet on fixed circular orbit in fully radiative discs at four different locations. The distance from star to planet changes (from top to bottom): $r_P=0.6$, $r_P=1.0$, $r_P=2.5$ and $r_P=4.0$ (in Jupiter radii). Please note the slightly different colour scale for each plot.
   \label{fig:Mig2DRho}
   }
\end{figure}

In \citet{2009A&A...506..971K} we explained in great detail with the help of 2D surface density plots how the torque acting on the planet is created in fully radiative discs. The torque acting on a $20 M_{Earth}$ planet on a circular orbit at $r_P= a_{Jup}$ in a fully radiative disc is positive, indicating outward migration. In Fig.~\ref{fig:Mig2DRho} the 2D surface density of the $20 M_{Earth}$ planets at $r=0.6, 1.0, 2.5, 4.0 a_{Jup}$ (from top to bottom) is displayed. The origin of the structure of the standard $r_P=1.0 a_{Jup}$ case (second from top) was described in \citet{2009A&A...506..971K}, and we display this case here again for better reference.

The planet located at $r_P=0.6 a_{Jup}$ (top panel) is still prone to outward migration (see positive torque in Fig.~\ref{fig:MigTorqueMass}), but at a slower rate. The surface density distribution shows some significant differences compared to the $r_P=1.0 a_{Jup}$ planet. The density increase ahead and inside of the planet ($\phi > 180^\circ$ and $r<0.6$) is still visible in the $r_P=0.6 a_{Jup}$ case, but the density decrease behind the planet ($\phi < 180^\circ$ and $r > 0.6$) is not that clearly visible. Indeed it seems as if the density behind the planet increased in a way that the total torque acting on the planet is reduced dramatically, resulting in reduced positive torque acting on the planet.

For the planet at $r_P = 2.5 a_{Jup}$, where the total torque acting on the planet is about zero, the density increase ahead of the planet is no longer visible in the surface density plot, but the decrease behind the planet is clearly visible. It also seems that the planet is able to deplete a larger area around it, possibly owing to the onset of gap formation. For even longer distances to the central star (e.g. $r_P = 4.0 a_{Jup}$), the effect becomes ever stronger, and the gap is more pronounced. The density increase in front of the planet is no longer visible at all. The torque acting on this planet is clearly negative, indicating inward migration. With increasing distance from the host star, the opening angle $\delta$ of the spiral arms becomes smaller (see Fig.~\ref{fig:Mig2DRho}), as can be inferred from $\delta \approx c_s/v_{Kep}$ which scales  $\propto r^{-0.35}$ for our temperature gradient of $T(r) \propto r^{-1.7}$.

To analyse the extent of outward migration from our disc properties in more detail, it seems useful to compare various time scales: the libration, the radiative and the viscous time scale in the disc. The necessary unsaturated torques needed for sustained outward migration require approximately equal libration and radiative diffusion times (see \citet{2008ApJ...672.1054B}; \citet{2008A&A...485..877P}). For the latter we use in our case the radiative diffusion time scale, $\tau_{rad}$. We define \citep{2010A&A.523...A30}

\begin{equation}
  	\tau_{rad}=\frac{s^2}{D_{rad}} \ ,
\end{equation} 
with the diffusion coefficient
\begin{equation}
	D_{rad}  =  \frac{4 c a T^3}{3 c_v \rho^2 \kappa} \ .
\end{equation}
For the typical diffusion length $s$ we substitute the vertical disc height, i.e. $s=H$, where
we use $H = c_s / \Omega$ with the sound speed $c_s$.  
The libration time given by \citep{2008ApJ...672.1054B}
\begin{equation}
  \tau_{lib} = 8 \pi r_{P} / (3 \Omega_K x_s) \ ,
\end{equation}
where $x_s$ denotes the half-width of the horseshoe-orbit, $x_s=1.16 r_P \sqrt{q / (\sqrt{\gamma}H/r)}$. 
Similarly to the radiative diffusion time, the viscous time scale is given by $\tau_{visc} = s^2/\nu$, again with $s=H$, and a constant $\nu$.
To compute the time scales all required quantities are evaluated in the midplane of the unperturbed disc at the beginning of the simulations. This applies to the density, temperature, opacity $\kappa(\rho,T)$, and the sound speed and $\Omega$.

\begin{figure}
 \centering
 \includegraphics[width=0.9\linwx]{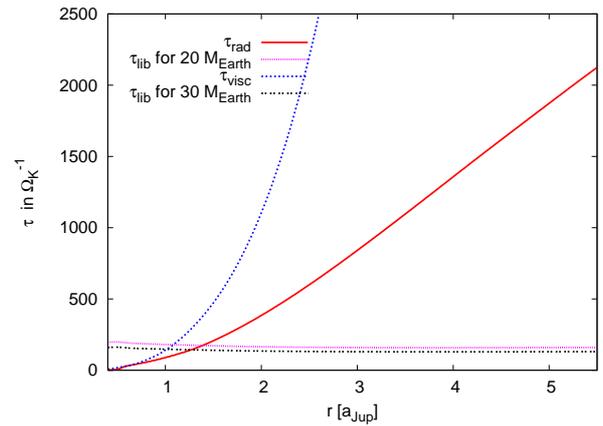}
 \caption{Radiative and viscous diffusion time scales that depend on the distance from the central star for our standard disc model. 
To compute the time scales we used the density and temperature of the midplane at the beginning of the fully radiative simulations (when the disc is in the $r - \theta$ equilibrium between viscous heating and radiative transport/cooling). Libration time scales are stated for $20$ and $30  M_{Earth}$ planets.
   \label{fig:Migtauall}
   }
\end{figure}

The three time scales are displayed in Fig.~\ref{fig:Migtauall}. For accretion discs that are solely heated internally by viscous dissipation we expect in equilibrium $\tau_{rad} \approx \tau_{visc}$. Apparently, this relation is fulfilled well (for $r<1.5 a_{Jup}$). We have plotted the libration time for two different planet masses, $20$ and $30 M_{Earth}$. Time scale arguments suggest a most efficient outward migration for equal $\tau_{lib}$ and $\tau_{rad}$, which is indeed roughly what we find in our 3D simulations. However, the overall range of outward migration is surprisingly broad. Specifically we find that $20 M_{Earth}$ planets are prone to outward migration up to about $r \approx 2.4$, where the two time scales differ by a factor of 3-4. For $30 M_{Earth}$ planets the range of outward migration is substantially smaller and centres directly around equal libration and radiative time scale.

We have checked the above estimates of the torques for a stationary planet with additional simulations of $20 M_{Earth}$ planets in discs, starting at $r=2.0 a_{Jup}$ and $r=3.0 a_{Jup}$ respectively, which were able to move freely inside the disc. The planets gather in this case indeed at the zero torque radius (results are not displayed here).

\section{Influence of the disc's mass}
\label{sec:discmass}

In this section we examine the influence of the disc's mass on the migration of low-mass planets embedded in these discs. First we compare the relevant physical properties of the discs with different masses, and then we investigate the planetary migration in those discs. We then finally discuss convection inside fully radiative discs.

\subsection{Properties of discs with different masses}

In our previous work, the disc's mass was fixed to $0.01 M_\odot$. We now extend the range of disc masses from $0.005 M_\odot$ to $0.04 M_\odot$ (with respect to the standard radial distance, from $0.4-2.5$). 
All models started locally isothermal with $H/r=0.05$ and during initial evolution on time this will
change to the appropriate equilibrium configurations (between viscous heating and radiative transport/cooling).

\begin{figure}
 \centering
 \includegraphics[width=0.9\linwx]{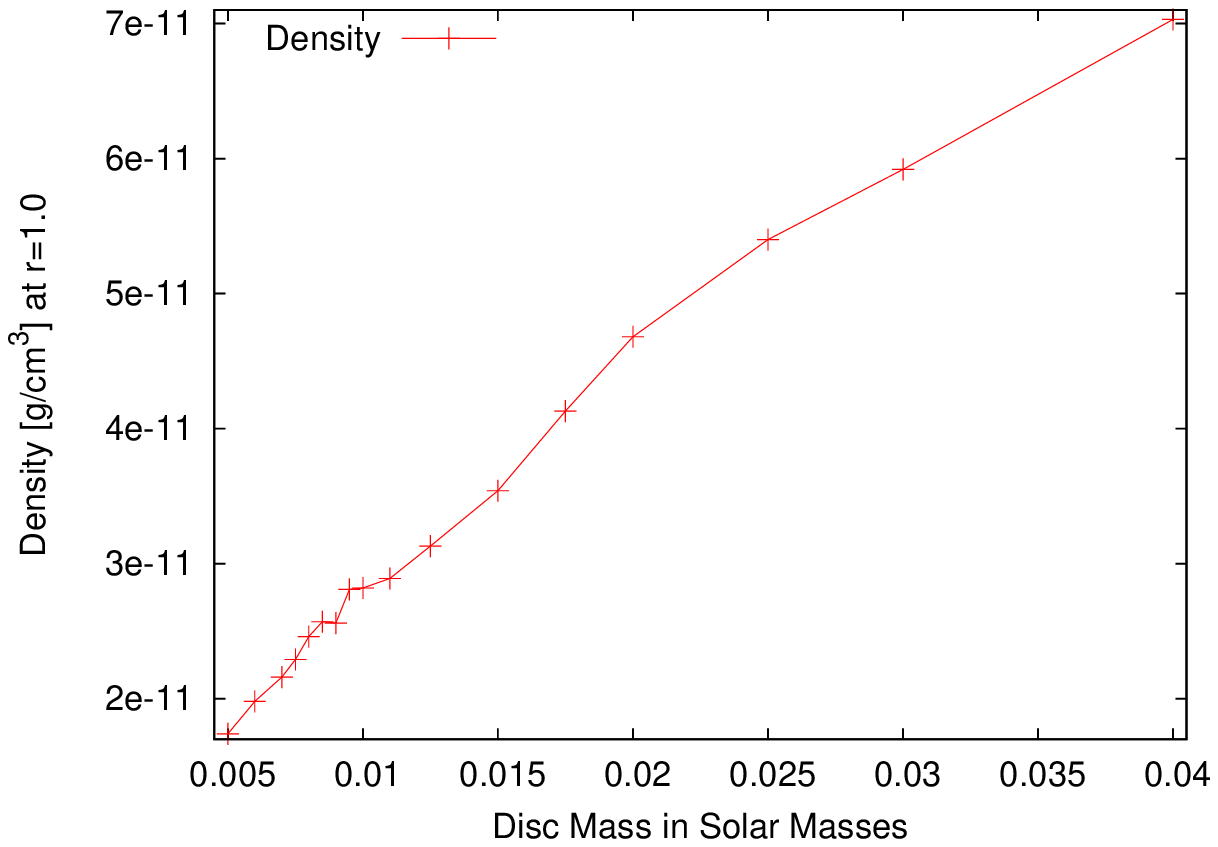}
 \includegraphics[width=0.9\linwx]{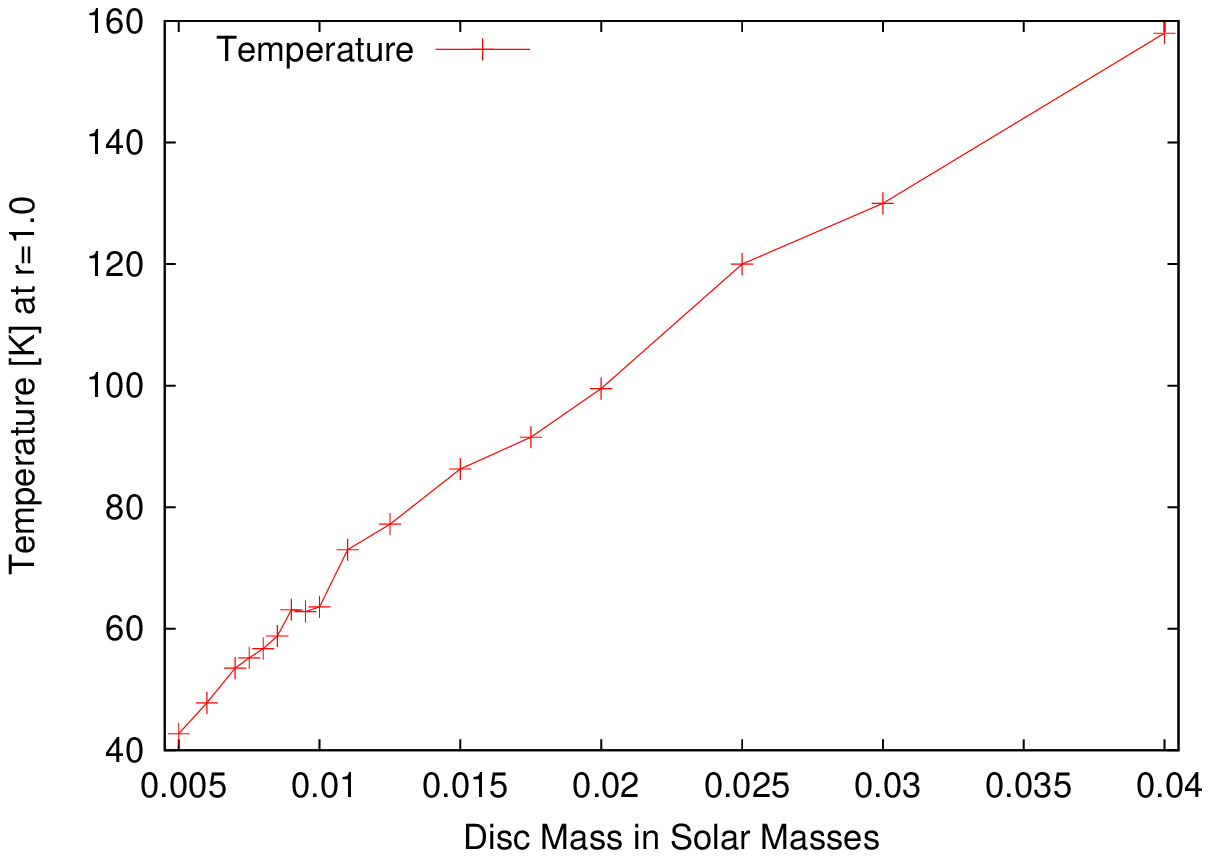}
 \includegraphics[width=0.9\linwx]{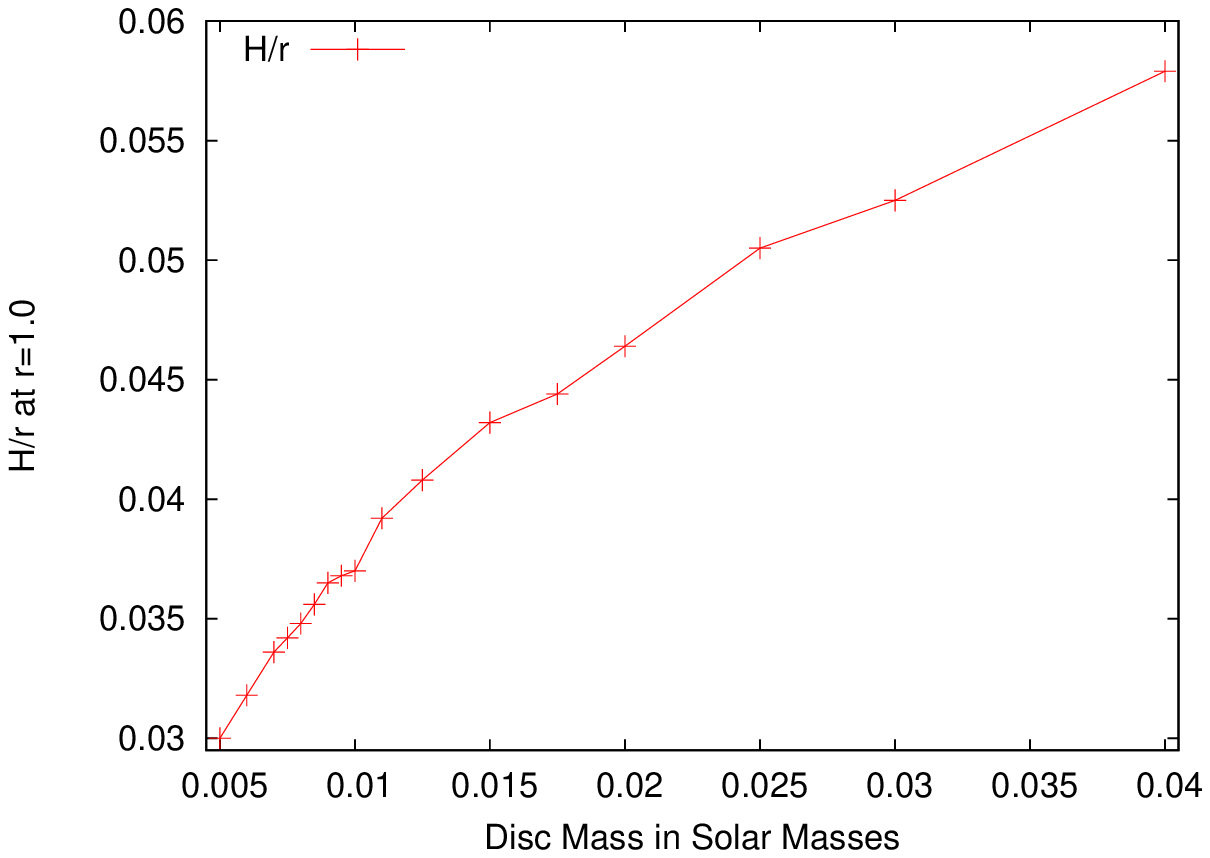}
 \caption{Density (top), temperature (middle) and aspect ratio (bottom) of discs with different masses in the initial equilibrium state. All quantities are measured in the disc's midplane at the reference distance, $r=1.0$.
   \label{fig:MassRTHR}
   }
\end{figure}

In Fig.~\ref{fig:MassRTHR} we display the density, temperature, and the aspect ratio of the equilibrium discs for different disc masses at $r=1.0$. Density and temperature, and $H/r$ are evaluated in the disc's midplane. The results are as expected from our previous simulations. A higher mass of the disc results in a higher density, temperature, and aspect ratio of the disc in the equilibrium state. In the isothermal case, a higher aspect ratio of the disc would result in a slower inward migration of a low-mass planet, see \citet{2002ApJ...565.1257T} for linear analysis and e.g. \citet{2010A&A.523...A30} for non-linear simulations. However, low-mass planets in fully radiative discs migrate outwards, so that the linear isothermal approach is not valid any more.

\begin{figure}
 \centering
 \includegraphics[width=0.9\linwx]{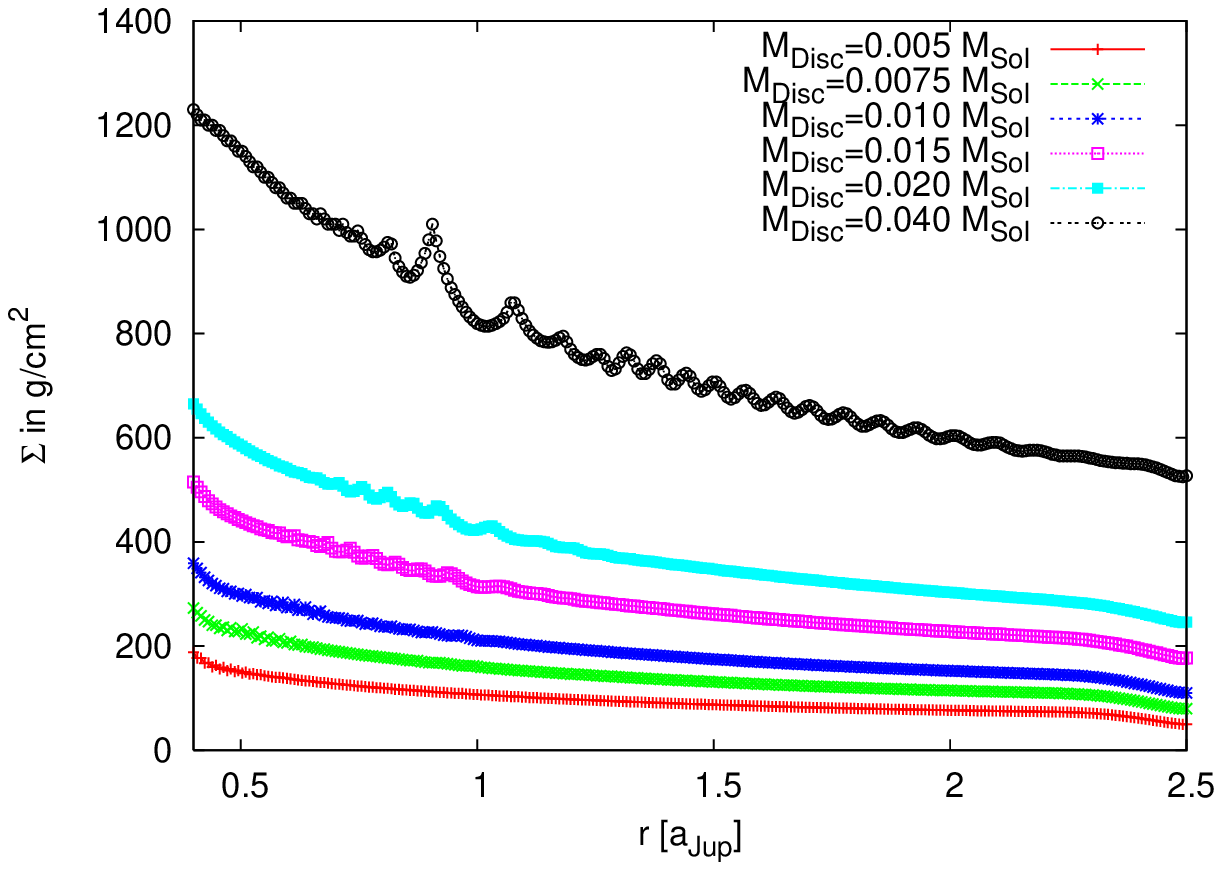}
 \includegraphics[width=0.9\linwx]{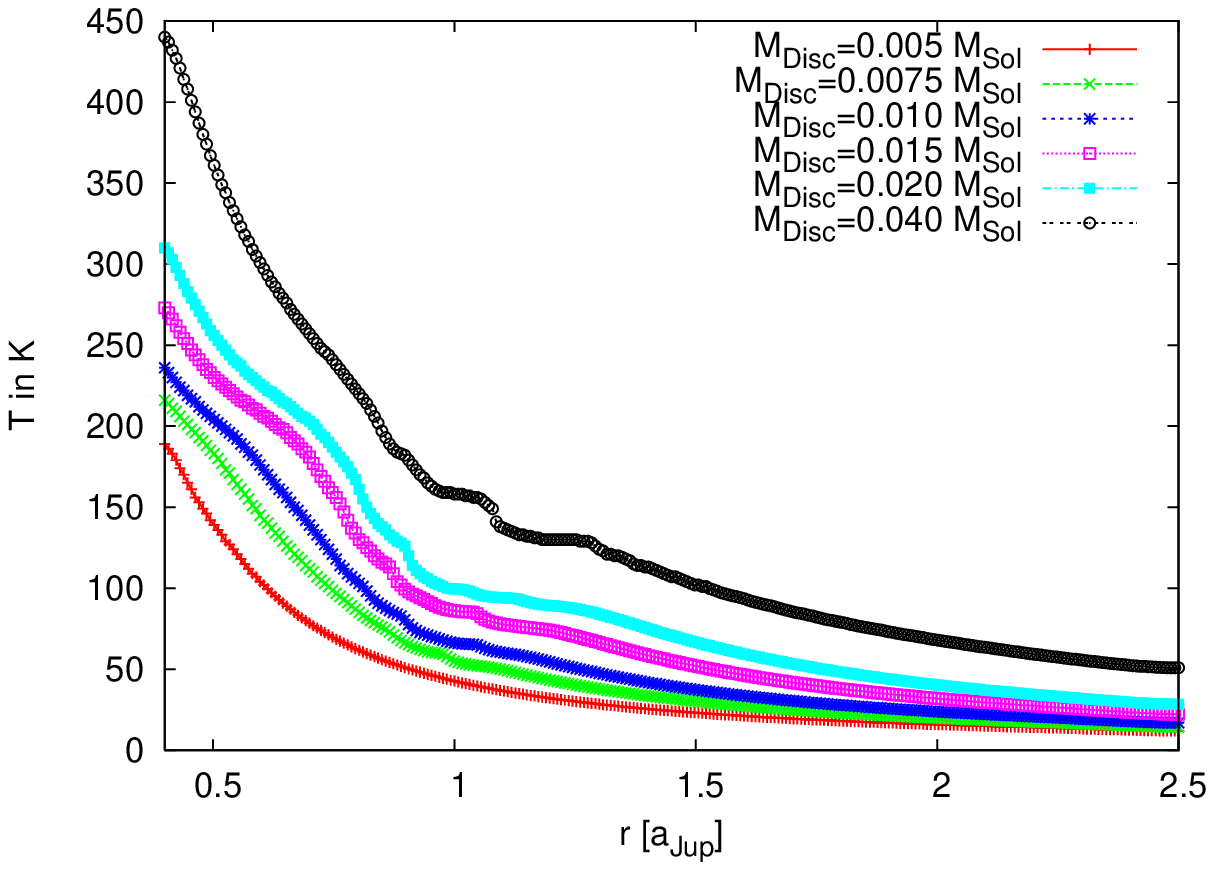}
 \caption{Surface density (top) and temperature (bottom) of discs with different masses in the equilibrium state of the disc. The temperature is measured in the disc's midplane.
   \label{fig:SigTempMass}
   }
\end{figure}

In Fig.~\ref{fig:SigTempMass} the radial distributions of surface density (top) and temperature (bottom) 
are displayed. The temperature has been measured in the disc's midplane. By construction, the surface density increases for higher disc masses, while it falls off with increasing distance to the star, on average
according to $\Sigma(r) \propto r^{-1/2}$ as expected for a  constant $\nu$. 
The surface density profiles for the higher mass discs with $M_{Disc} \geq 0.015 M_\odot$ show some fluctuations. With increasing disc mass, these fluctuations become stronger and reach out to a longer distance from the star. While they are quite short for $M_{Disc} \leq 0.02 M_\odot$ and reach only to $r \approx 1.3$, they become very strong and reach out to $r \approx 2.3$ for $M_{Disc}=0.04 M_\odot$. These fluctuations of the surface density vary in time and are related through convective motions in the disc, see below.

The described fluctuations in the surface density can also be seen in the temperature profiles of discs with different disc masses (bottom panel in Fig.~\ref{fig:SigTempMass}). The variabilities of the temperature are not as strong as those for the surface density, nevertheless, they are clearly notable for $M_{Disc} \geq 0.015 M_\odot$ and increase with the disc's mass. They also, change in time, as does the surface density. A higher disc mass seems to support stronger fluctuations that reach farther out into the disc. 

Because a higher disc mass results in a higher aspect ratio of the disc, the cut-off of the computed domain (at $7^\circ$ above and below the midplane) might change the structure of the disc. Additional simulations with larger $\theta$ did not change the density and temperature patterns at $\theta=83^\circ$ and $\theta=97^\circ$ for all disc masses \citep[see also][]{2009A&A...506..971K}. However, a too low boundary substantially changes the distributions and might therefore influence convection in the disc as well.

The changes in the surface density profiles have a direct influence on the migration of an embedded planet. If a planet is embedded in a region in the disc where the fluctuations of the surface density are very strong (i.e. strong convection), the direction of migration might not be clearly determinable, because changes in the surface density profile directly influence the migration. Stationary gradients in surface density profiles can even be used as planetary traps to collect planetary embryos \citep{2008A&A...478..929M}.

\subsection{Influences on the migration of low-mass planets}

As mentioned above, a different aspect ratio of the disc will change the rate of planetary migration. In the isothermal case, a higher aspect ratio will result in slower inward migration in theory \citep{2002ApJ...565.1257T}, which we supported in our previous simulations \citep{2010A&A.523...A30, 2011A&ABitschKley}. A higher disc mass results in a higher aspect ratio and temperature of the disc, and we may expect changes in the migration rates. When the planet is farther away from the central star, the temperature and density of the disc are reduced. If the reduction in density and temperature is sufficient, the planets stop their outward migration (see Fig.~\ref{fig:MigTorqueMass}). One might now expect that for discs with very low masses the torque acting on a $20 M_{Earth}$ planet at $r=1.0 a_{Jup}$ might become negative. Very high temperatures and high densities inside the disc, on the other hand, might influence the outward migration as well. Following the formula of \c
 itet{2010MNRAS.401.1950P} in Equation (\ref{eq:paar09}), one might suspect stronger torques acting on planets in more massive discs for constant $\alpha$, $\beta$ and $\xi$ (increase in surface density overcompensates the increase in aspect ratio).

In the top panel of Fig.~\ref{fig:MassTorque} the total torque acting on $20 M_{Earth}$ planets on circular orbits embedded in fully radiative discs with different masses and the theoretical results from \citet{2010MNRAS.401.1950P} and \citet{2011MNRAS.410..293P} are displayed (blue and purple). The torque  acting on the planet remains nearly constant within a small interval around our standard disc mass of $0.01 M_\odot$. For lower disc masses, the torque drops off very rapidly to even negative values for $M_{Disc}=0.005 M_\odot$, as we expected. For higher disc masses, the torque drops off as well. First at a faster rate (to $\approx 0.020 M_\odot$), then at a slightly slower rate, until it reaches an about zero torque state for $M_{Disc}=0.040 M_\odot$. This contradicts to our first expectation that planets in more massive discs should experience a higher torque, the reason may be a change in the temperature gradient and the influence of convection.

When looking at the surface density profile displayed in Fig.~\ref{fig:SigTempMass}, it is clear that the changes in the surface density may disrupt the very sensitive density pattern near the planet. As the convection cells in the disc change with time, the torque acting on the planet will change as well, giving rise to high fluctuations/oscillations in the total torque acting on the planet.  Hence, the torques acting on the planet have been averaged over $20$ planetary orbits. After averaging, the torques acting on planets in convective discs show only very low fluctuations.

For disc masses around $\approx 0.01 M_\odot$ the theoretical formula from \citet{2010MNRAS.401.1950P} (see eq.~\ref{eq:paar09}) fits our 3D simulations up to a factor of $25\%$. For very low disc masses ($M_{disc}=0.005 M_\odot$), however, the fit is not as good. This may be because of the reduced disc mass and the consequently changed surface density distribution (which changes the torque acting on the planet), as explained in Section \ref{sec:outwardrange}. For higher disc masses ($M_{disc} \approx 0.015 M_\odot$), the two torque values differ more and more. As the torques of our simulations tend to go to zero, the theoretically predicted adiabatic torques from \citet{2010MNRAS.401.1950P} become even more extended.

The formula from \citet{2011MNRAS.410..293P}, which includes the important effects of viscosity and heat diffusion, differs by a factor of three near disc masses of $M_{disc} \approx 0.01 M_\odot$ as one could have expected from the results presented in Fig.~\ref{fig:MigTorquePaar}. For higher disc masses, the torques from \citet{2011MNRAS.410..293P} remain nearly unchanged. The torques from our simulations are reduced, but they do not reach negative values. Besides the differences described in Section \ref{sec:outwardrange}, more differences arise from convection in the disc, because convection results in fluctuations in the torque, which is not considered in the analytical formulae. However, convection seems to play a role only for discs with $M_{disc} > 0.02 M_\odot$.

\subsection{Torque analysis}

\begin{figure}
 \centering
 \includegraphics[width=0.9\linwx]{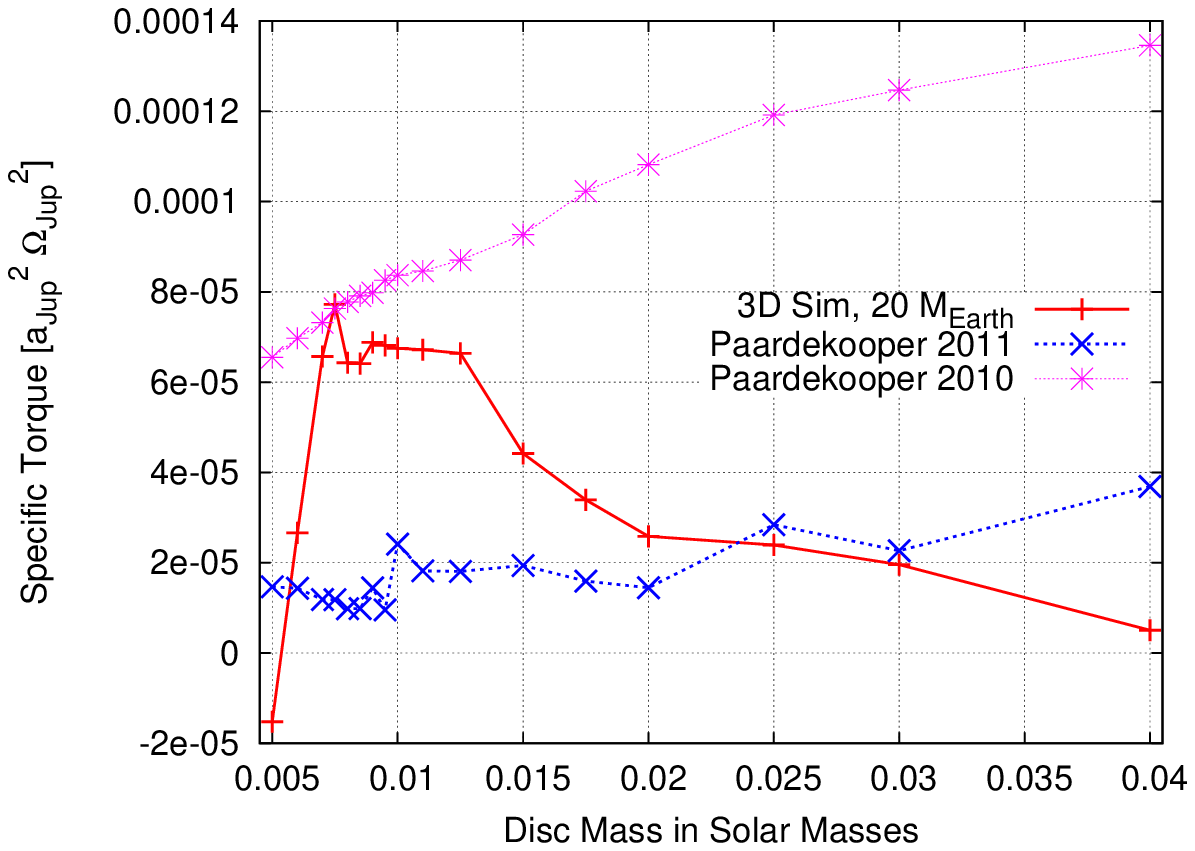}
 \includegraphics[width=0.9\linwx]{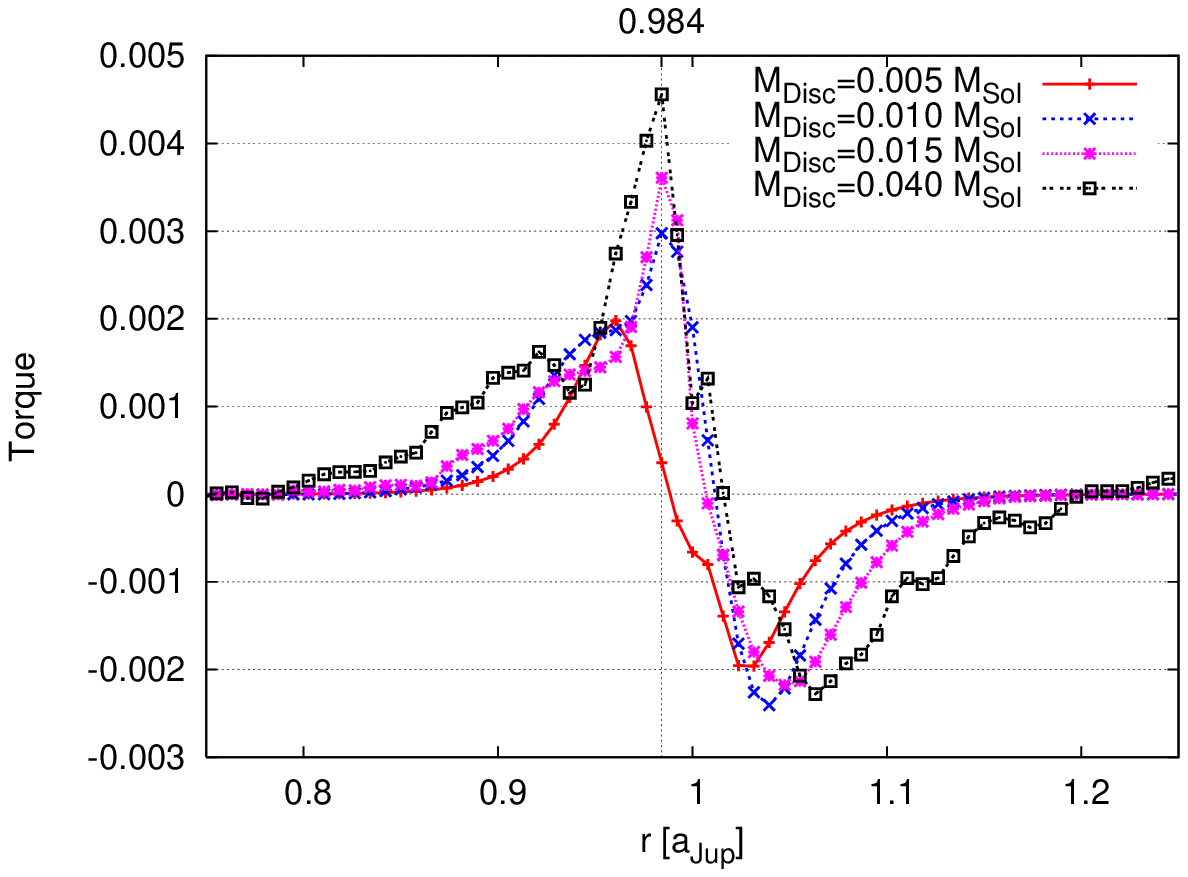}
 \caption{Torque acting on a planet located at 5 AU for different disc masses.
{\bf Top:} Specific total torque acting on planets ($20 M_{Earth}$) embedded in discs with different disc masses. The planets embedded in the higher mass discs $M_{Disc} > 0.02 M_\odot$ are in the convective zone in the disc, so that the torque acting on the planet is very noisy and has been averaged over 20 planetary orbits. Additionally, we over-plotted results (blue and purple) from the theoretical torque formulae of \citet{2010MNRAS.401.1950P, 2011MNRAS.410..293P} for $20 M_{Earth}$ planets. {\bf Bottom:} Radial torque density $\Gamma (r)$ acting on the planet for different disc masses. For comparison, the vertical line indicates the location of the maximum as found for our standard case.
   \label{fig:MassTorque}
   }
\end{figure}

In the bottom panel of Fig.~\ref{fig:MassTorque} the radial torque density $\Gamma (r)$ is displayed for different disc masses. For the lowest disc mass in our simulations, $M_{Disc}=0.005 M_\odot$, the usual spike in the torque density cannot be seen. The spike in the torque density distribution is an indication for a positive torque in a fully radiative disc \citep{2009A&A...506..971K}. For higher disc masses (up to $M_{Disc} \approx 0.02 M_\odot$), it is is clearly visible. For those disc masses, the total torque is indeed positive (see top panel of Fig.~\ref{fig:MassTorque}), indicating outward migration of the embedded protoplanet. 

\begin{figure}
 \centering
 \includegraphics[width=0.9\linwx]{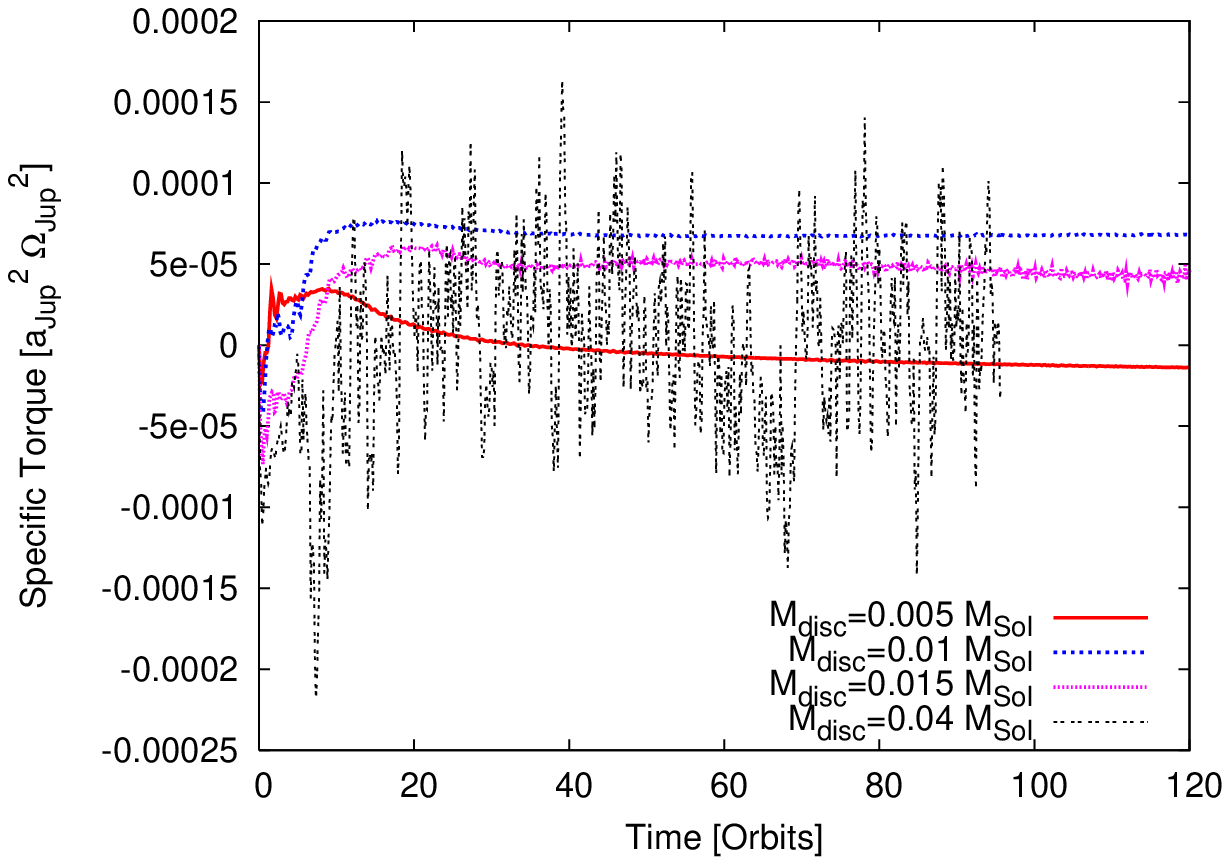}
 \includegraphics[width=0.9\linwx]{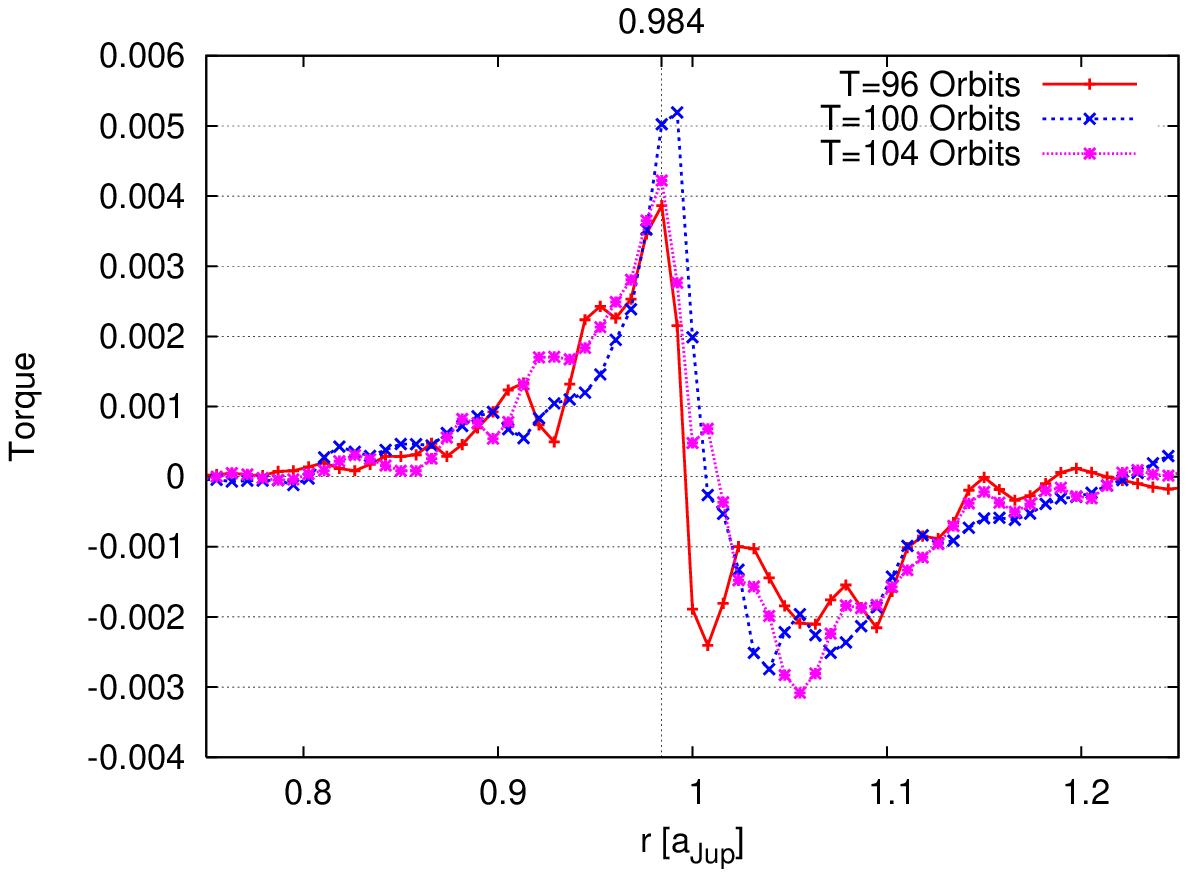}
 \caption{{\bf Top:} Specific total torque evolution acting on planets ($20 M_{Earth}$) embedded in discs with different disc masses. The planets embedded in the higher mass discs $M_{Disc} > 0.02 M_\odot$ are in the convective zone in the disc, so that the torque acting on the planet is very noisy compared to the low-mass discs. {\bf Bottom:} Radial torque density $\Gamma (r)$ acting on the planet embedded in a disc with $M_{disc}=0.04M_\odot$ at different stages of the evolution.
   \label{fig:TorqueM400}
   }
\end{figure}

The torque density for the $M_{Disc}=0.04 M_\odot$ disc seems to indicate a total positive torque acting on the planet, and it is
indeed positive at the moment of the snapshot, but as the fluctuations in the surface density change in time, so does the torque acting on an embedded planet. Therefore the torque density for the $M_{Disc}=0.04 M_\odot$ disc at one single moment during the evolution does not necessarily reflect the longterm outcome, if the fluctuations are to strong. This time variation of the total torque and torque density acting on the planet is displayed in  Fig.~\ref{fig:TorqueM400}. The top panel in Fig.~\ref{fig:TorqueM400} shows the time evolution of the total torque acting on embedded planets for discs with different disc masses. For $M_{disc} < 0.02 M_\odot$ the torque acting on the planet is constant in time (after about $50$ orbits), while it shows very high fluctuations for $M_{disc} = 0.04 M_\odot$. In the bottom panel of Fig.~\ref{fig:TorqueM400} the torque density $\Gamma (r)$ for a $20 M_{Earth}$ planet in a $M_{disc} = 0.04 M_\odot$ disc is shown. The torque density is plott
 ed at different times. Because the total torque was fluctuating very much, it is no surprise to find these fluctuations for the torque density as well. These fluctuations, induced by convection, clearly show that a higher disc mass disturbs the evolution of the torque.

\begin{figure}
 \centering
 \includegraphics[width=0.85\linwx]{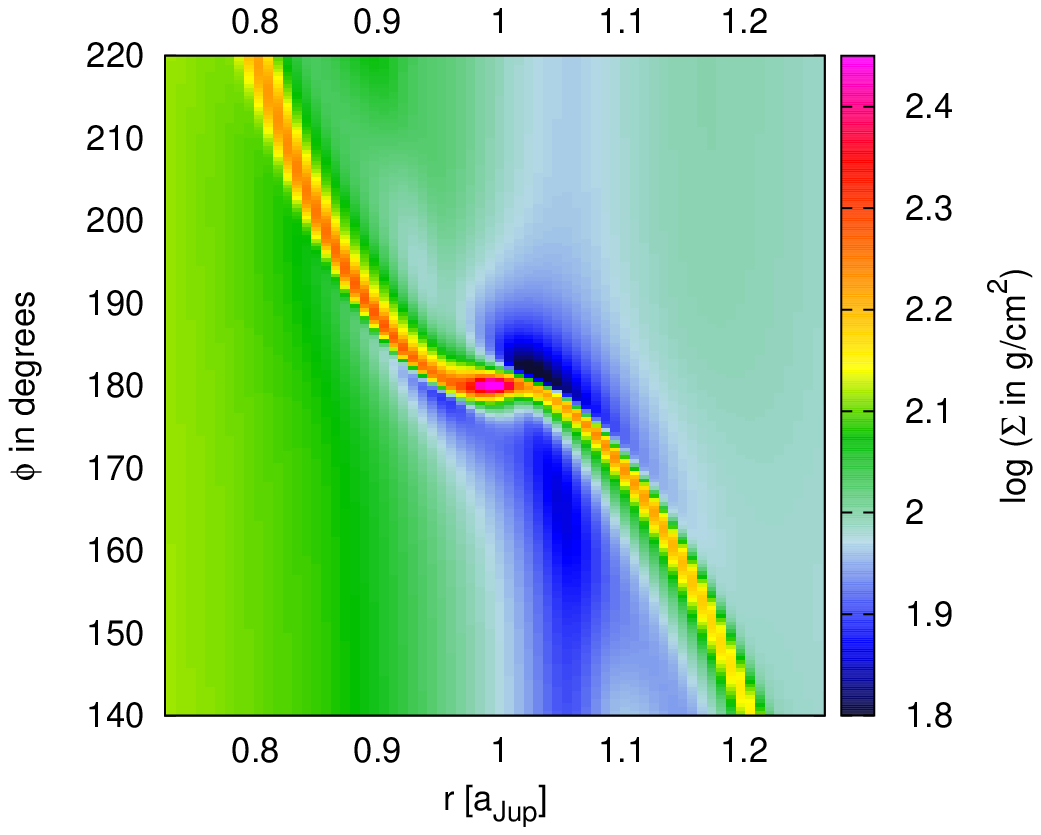}
 \includegraphics[width=0.85\linwx]{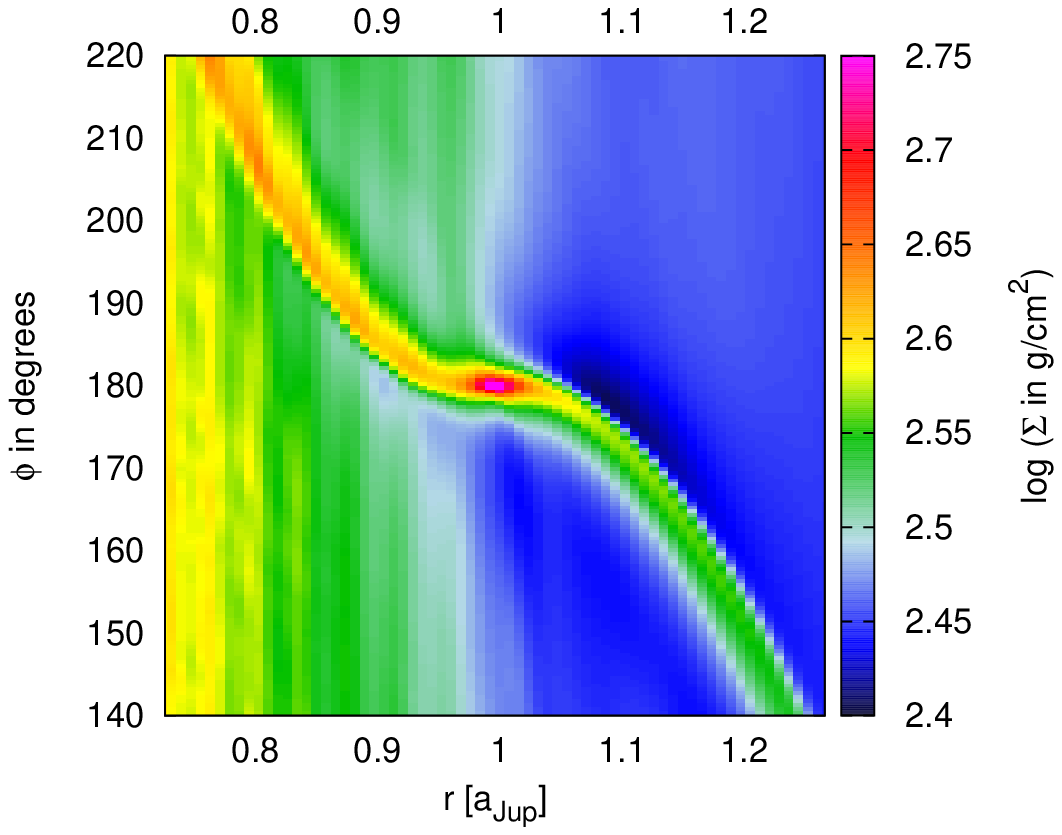}
 \includegraphics[width=0.85\linwx]{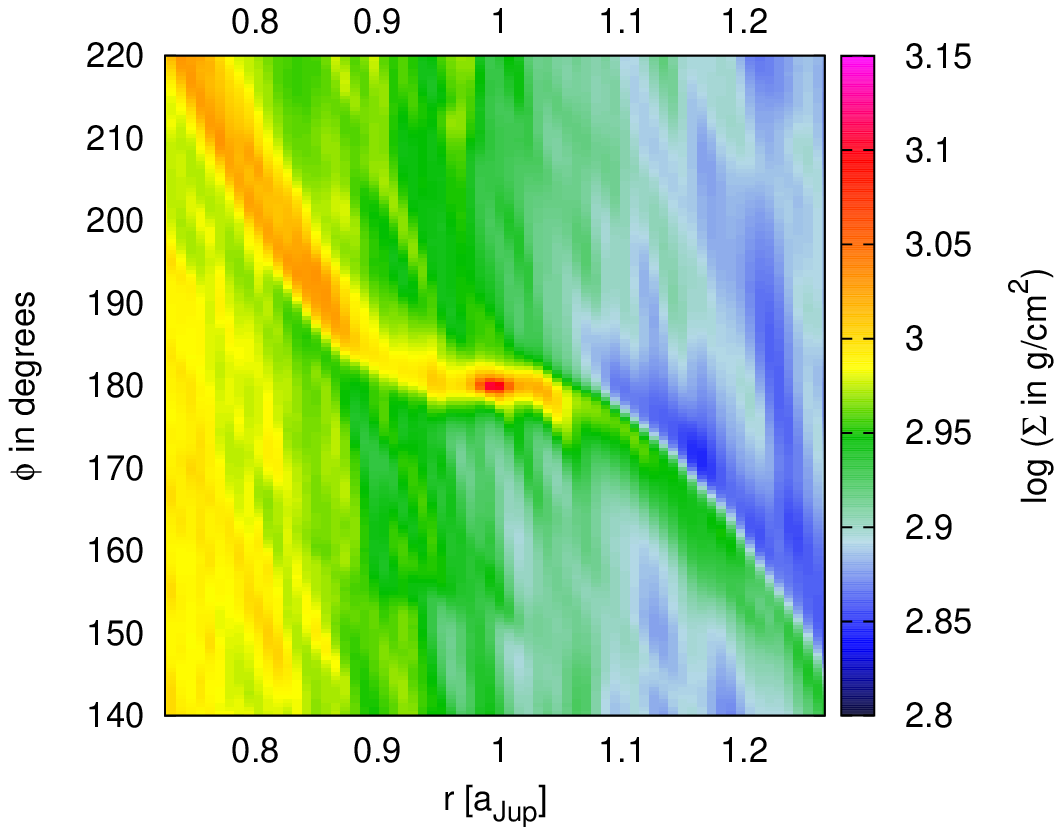}
 \caption{Surface density maps for planets on fixed circular orbits in fully radiative discs  with different disc masses (from top to bottom): $M_{Disc} = 0.005 M_\odot$, $M_{Disc} = 0.015 M_\odot$ and $M_{Disc} = 0.040 M_\odot$. The disruption in the surface density patterns for the higher mass discs are caused by convection inside the disc.
   \label{fig:Mass2DRho}
   }
\end{figure}

In Fig.~\ref{fig:Mass2DRho} the surface density profiles of fully radiative discs with disc masses of $0.005$, $0.015$ and $0.04 M_\odot$ with embedded $20 M_{Earth}$ planets are displayed. The planet in the $0.005 M_\odot$ disc generates a very similar surface density structure compared to our standard $0.01 M_\odot$ disc (second panel in Fig.~\ref{fig:Mig2DRho}). The overall density is, of course, reduced (because the disc mass is much lower), but the general pattern of the density increases in front of the planet $\phi > 180^\circ$ and $r < 1.0$ and the decrease behind the planet $\phi < 180^\circ$ and $r > 1.0$ remains intact. However, the density structure relevant for outward migration is not as pronounced as it should be to result in a positive torque acting on the planet. 

For planets embedded in discs with higher masses, the picture is quite different. For a disc mass of $0.015 M_\odot$ (middle picture in Fig.~\ref{fig:Mass2DRho}) the density structure in the direction of the star ($r < 0.9 a_{Jup}$) is very distorted, but one can still see the density increase ahead and the density decrease behind the planet. The distortion seems to reduce the torque acting on the planet, but the overall torque is still positive, indicating outward migration. For an even higher disc mass (bottom picture in Fig.~\ref{fig:Mass2DRho} with $M_{Disc}=0.040 M_\odot$) the distortions in the disc increase more. The density structure, normally seen for low-mass planets in fully radiative discs, is no longer visible at all. The distortions are so strong that the torque acting on the planet becomes about zero, indicating only a low migration rate.

However, in this last case the torque is in a state of constant fluctuations, which complicates realistic predictions about the direction of migration in these massive discs. The fluctuations of the torque have their cause in fluctuations of the density patterns, which indicates that the convective zone inside the disc is enlarged compared to low-massive discs. We observed the phenomenon of convection briefly in our previous work \citep{2009A&A...506..971K} for discs with $M_{Disc}=0.01 M_\odot$ as well, but the convective zone did not reach the planet, and thus did not disturb the density pattern around the planet.

\begin{figure}
 \centering
 \includegraphics[width=0.9\linwx]{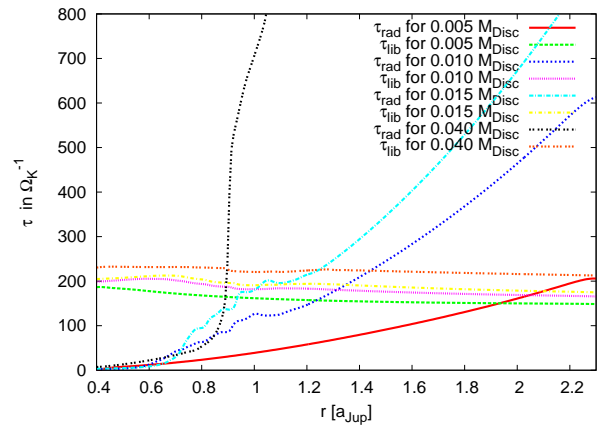}
 \caption{Time scales depending on  the distance from the central star for $0.005 M_\odot$, $0.010 M_\odot$, $0.015 M_\odot$ and $0.040 M_\odot$ discs. To compute the time scales we used the density and temperature of the midplane at the beginning of the fully radiative simulations (when the disc is in the $r - \theta$ equilibrium between viscous heating and radiative transport/cooling).
   \label{fig:Masstauall}
   }
\end{figure}

In Fig.~\ref{fig:Masstauall} we display the radiative diffusion time scale and the libration time scale for discs with different masses. In order to keep the torques acting on the planet unsaturated (to evoke outward migration), the libration and radiative diffusion time scale need to be approximately equal \citep{2008ApJ...672.1054B, 2008A&A...485..877P}. However, for convective discs the equilibrium state through cooling through convection is reached only when $\tau_{rad} > \tau_{visc}$ as observed in Fig.~\ref{fig:Masstauall}. The time scales for the $0.005 M_\odot$ disc indicate outward migration in a region at $r \approx 2.0 a_{Jup}$, but at $r=1.0 a_{Jup}$ the time scales differ by a factor of $4$, indicating that the torques are not kept unsaturated in this region of the disc, which in turn indicates inward migration (as presented in in the top figure in Fig.~\ref{fig:MassTorque}). 

For the $0.015 M_\odot$ disc the time scales are nearly identical at $r=1.0 a_{Jup}$, which indicates outward migration (as can be seen in the top of Fig.~\ref{fig:MassTorque}). However, for longer distances to the central star, the time scales start to differ, which indicates inward migration. In the $0.040 M_\odot$ disc, the time scales differ by a factor of $3$ at $r=1.0 a_{Jup}$, which indicates inward migration. However, the measured torque acting on the planet is positive, indicating slow outward migration. But because the planet is embedded in a highly convective region in the disc, it is very difficult to predict the motion of the planet correctly by considering only the time scales.

\subsection{Orbital evolution}

\begin{figure}
 \centering
 \includegraphics[width=0.9\linwx]{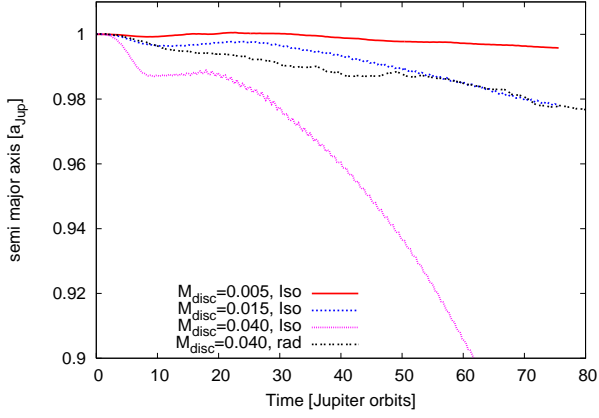}
 \caption{Evolution of semi-major axis for $20 M_{Earth}$ planets (starting at $r=1.0 a_{Jup}$) in isothermal and fully radiative discs with different disc masses.
   \label{fig:Mdiscmove}
   }
\end{figure}

In Fig.~\ref{fig:Mdiscmove} the evolution of semi-major axis for $20 M_{Earth}$ planets in isothermal and fully radiative discs with different disc masses is displayed. The isothermal reference simulations are started from a $H/r=0.037$ disc, which represents the $H/r$ value at the planets starting location in the fully radiative regime. In the isothermal disc, no convection is present and therefore the embedded planet migrates as expected. A higher disc mass results in a faster inward migration. However, it seems that for the $M_{disc}=0.04 M_\odot$ disc the type-III-migration regime is hit, because the planet moves inwards very fast.

For a planet on a fixed circular orbit in a fully radiative disc with $M_{disc}=0.04 M_\odot$ we determined a positive torque (see Fig.~\ref{fig:MassTorque}) by averaging in time. However, the total torque acting on the planet undergoes strong fluctuations in time (see Fig.~\ref{fig:TorqueM400}). When embedding a $20 M_{Earth}$ planet in such a highly convective disc, the evolution pattern should to some extend reflect the strong fluctuations in the torque acting on a planet on a fixed orbit. Indeed, the planet experiences some small kicks in its evolution pattern (see Fig.~\ref{fig:Mdiscmove}). Interestingly, the planet migrates inwards despite the positive torque acting on a fixed planet. It seems that in the convective region of high-mass discs, the measurement of the torque for planets on fixed orbits is not as reliable as for planets in low-mass discs. Additionally, the time averaging may have to be performed over a longer time span.

\subsection{Convective zone}

\begin{figure}
 \centering
 \includegraphics[width=0.805\linwx]{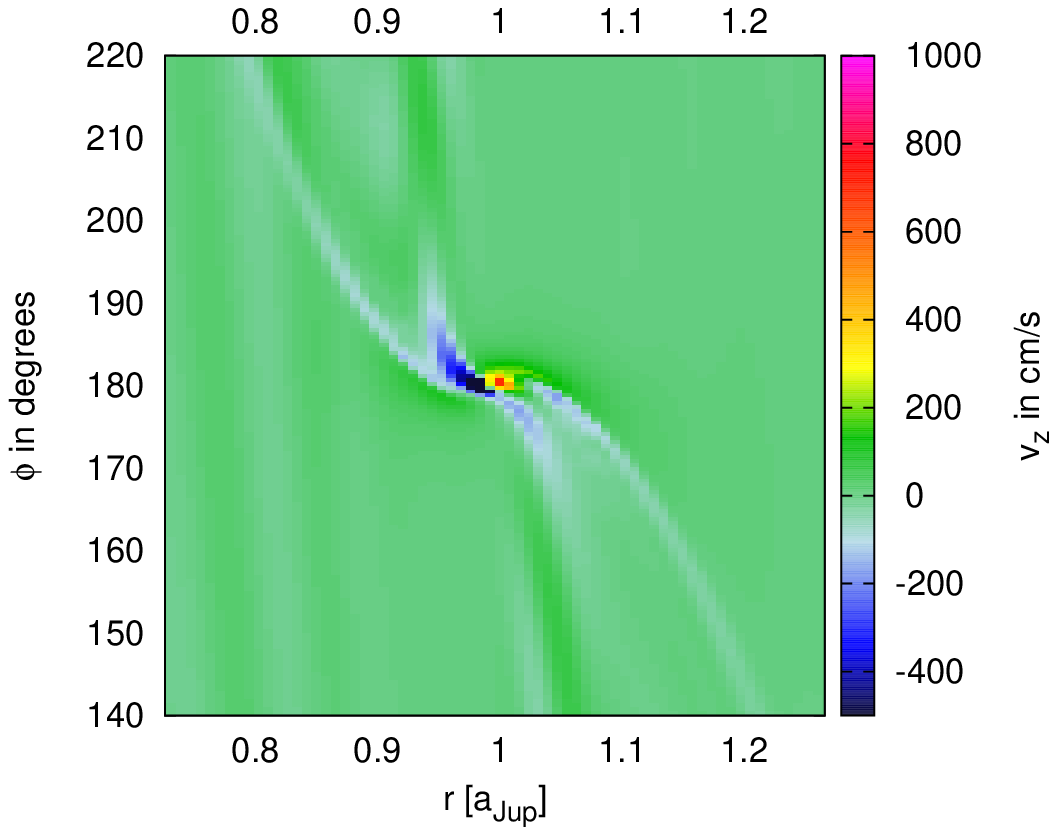}
 \includegraphics[width=0.805\linwx]{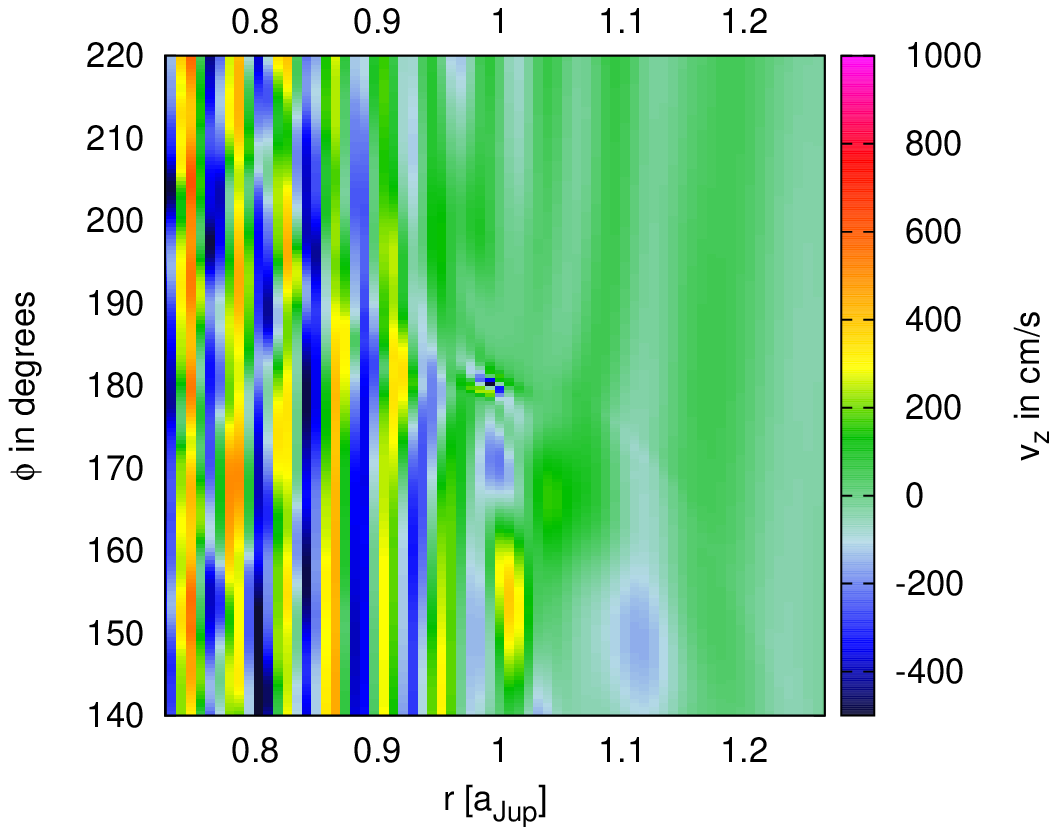}
 \includegraphics[width=0.805\linwx]{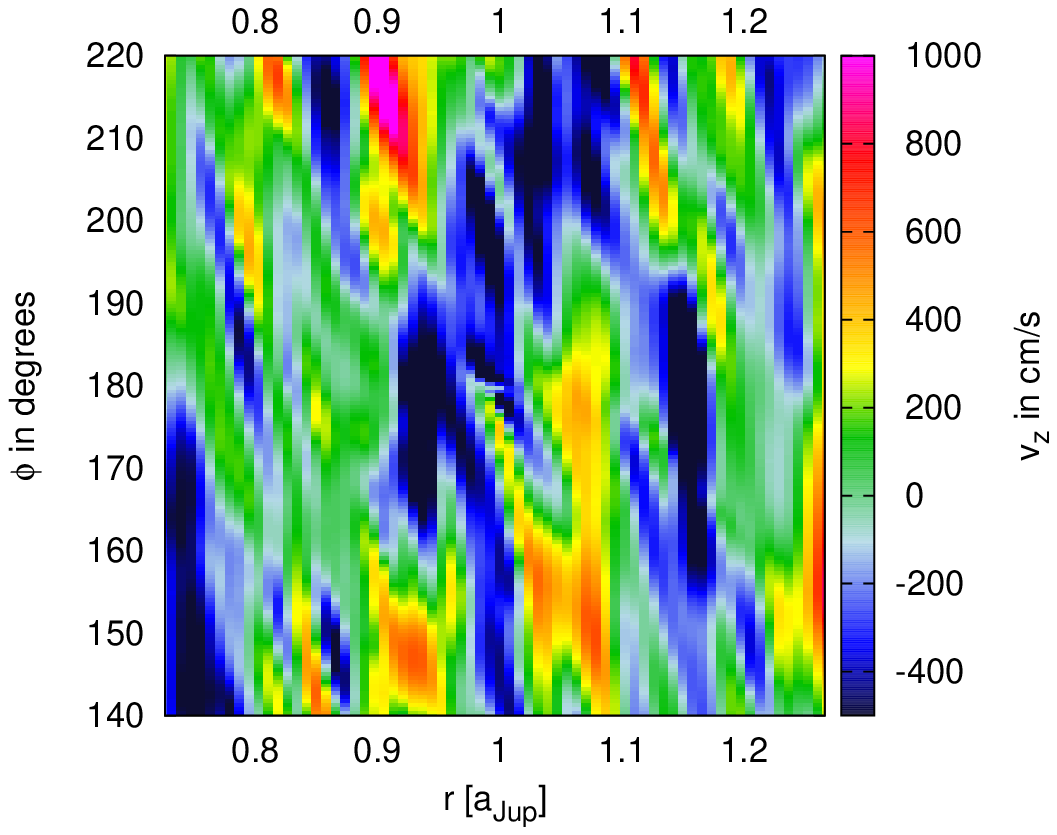}
 \caption{Vertical velocities in $z$-direction for planets on fixed circular orbits in fully radiative discs  with different disc masses (from top to bottom) in the disc's midplane: $M_{Disc} = 0.005 M_\odot$, $M_{Disc} = 0.015 M_\odot$ and $M_{Disc} = 0.04 M_\odot$. The disruption in the velocity patterns for the higher mass discs are caused by convection inside the disc. A positive velocity simulates outward flows, while a negative velocity indicates a flow towards the disc's midplane.
   \label{fig:Mass2Dvy}
   }
\end{figure}

In order to investigate whether convection is actually present in the disc, we display the velocities in $z$-direction (out of midplane of the disc) for different disc masses ($M_{Disc} = 0.005 M_\odot$, $M_{Disc} = 0.015 M_\odot$ and $M_{Disc} = 0.04 M_\odot$) in the disc's midplane in Fig.~\ref{fig:Mass2Dvy}. These plots represent simulations with a whole disc, meaning $83^\circ \leq \theta \leq 97^\circ$. As the surface density plots indicated, there are no disrupted areas in the velocity patters for the $M_{Disc} = 0.005 M_\odot$ simulations. Therefore for this low disc mass we observe no convection near the planet. 

However, for the $M_{Disc} = 0.015 M_\odot$ case the surface density patterns already indicated that convection is possible in the disc inside of the planet's distance to the central star. The velocity distributions confirm this result, lines with positive velocities (indicating a flow towards the upper boundary of the disc) alternate with lines with negative velocities (indicating a flow towards the midplane of the disc). These flows are a clear indicator of motion caused by convection in the disc. In the $M_{Disc} = 0.040 M_\odot$ case, the fluctuations in the surface density increased dramatically, as did the changes in the velocity pattern. Alternations between positive and negative velocities have increased and indicate a very strong convective region that disturbs the torque acting on the planet (and therefore it's migration). The flow pattern in the disc is very erratic, making it absolutely necessary to average the torque acting on an embedded planet for many orbits.

For simulations that cover only the upper half of the disc, the convection cells inside the disc end at the disc's midplane, but in reality these convective cells continue to the lower half of the disc. However, that convection inside the disc changes the behaviour of the planet disc interactions can also be seen for simulations containing only one half of the disc.

\begin{figure}
 \centering
 \includegraphics[width=0.805\linwx]{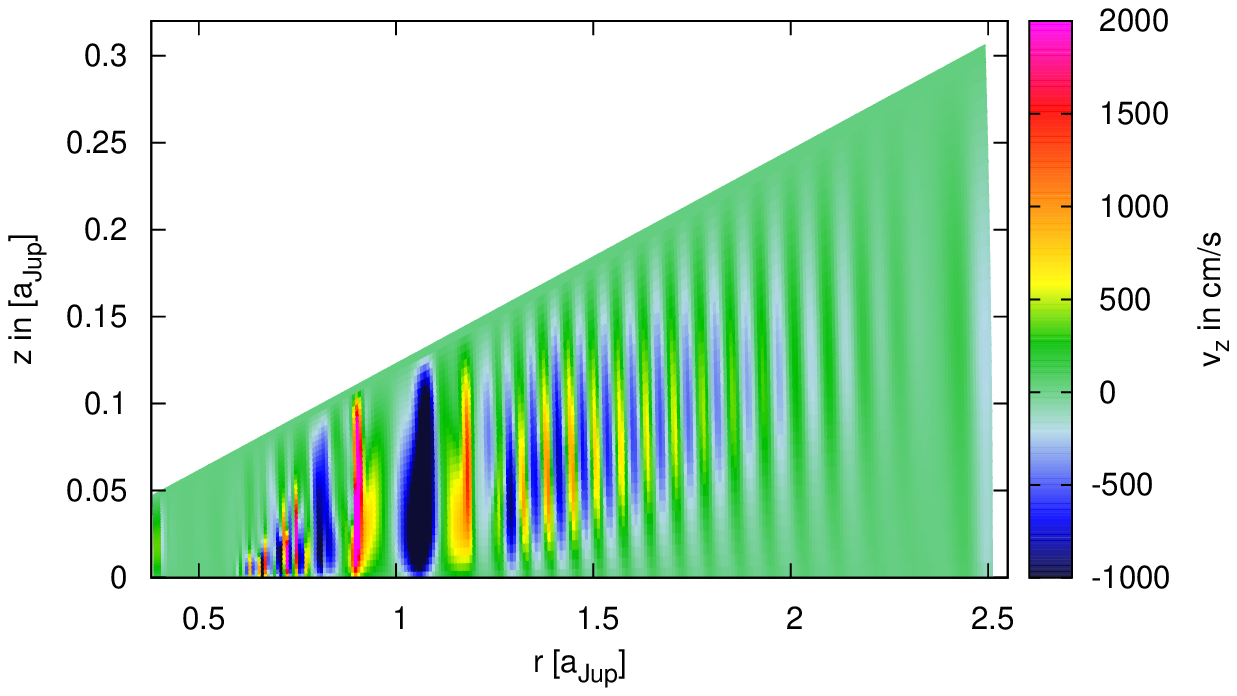}
 \includegraphics[width=0.805\linwx]{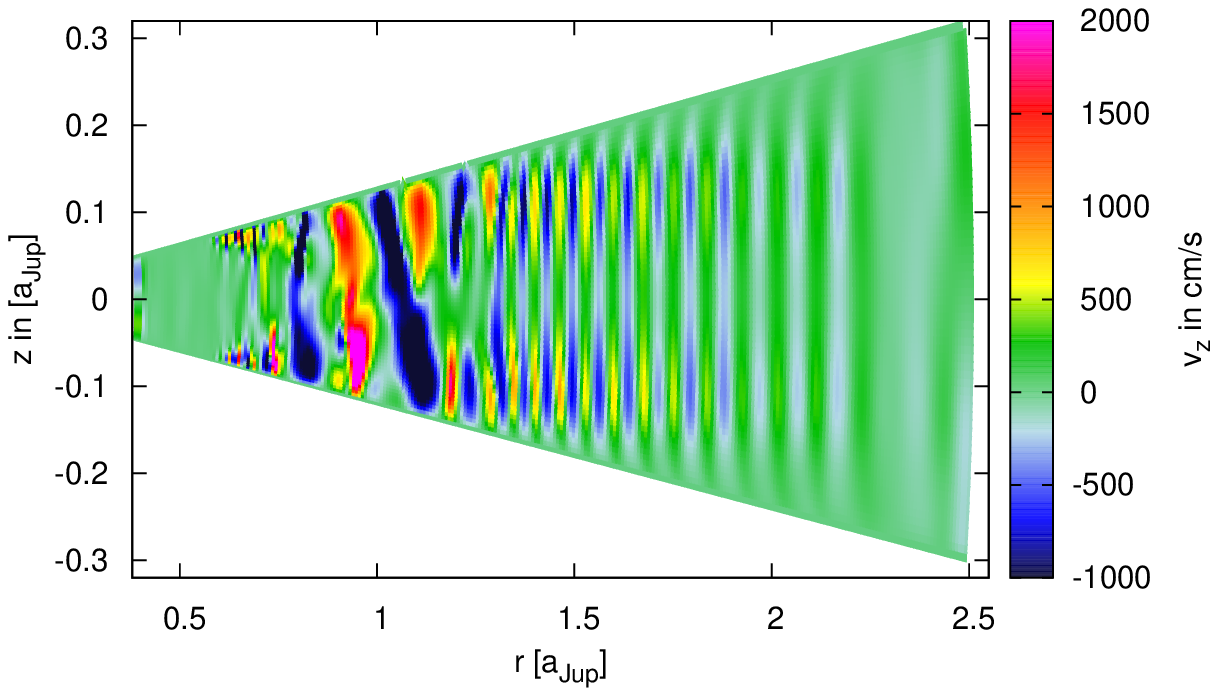}
 \caption{Velocities in $z$-direction for $0.04 M_\odot$ discs without planet. In the top panel only the upper half of the disc was simulated, while in the bottom panel both sides of the disc were simulated (with twice the number of grid cells in $\theta$ direction). The other simulation parameters are identical.
   \label{fig:Mass2Dthetavy}
   }
\end{figure}

In Fig.~\ref{fig:Mass2Dthetavy} the velocity in $z$ direction is displayed for a half-size disc (only the upper half of the disc is computed, top figure) and for a full disc. These computations have been performed only for fully radiative axisymmetric 2D discs (in $r$-$\theta$ direction) with a disc mass of $M_{disc}=0.04 M_\odot$ without an embedded planet. The convection in the disc is clearly visible. In both simulations the convection cells in the disc become more symmetric for distances longer than $r > 1.25 a_{Jup}$, meaning that the velocity changes from positive to negative only in radial direction and not in the vertical direction as well. For shorter distances to the central star the convection cells are very irregular in both cases.

\begin{figure}
 \centering
 \includegraphics[width=0.805\linwx]{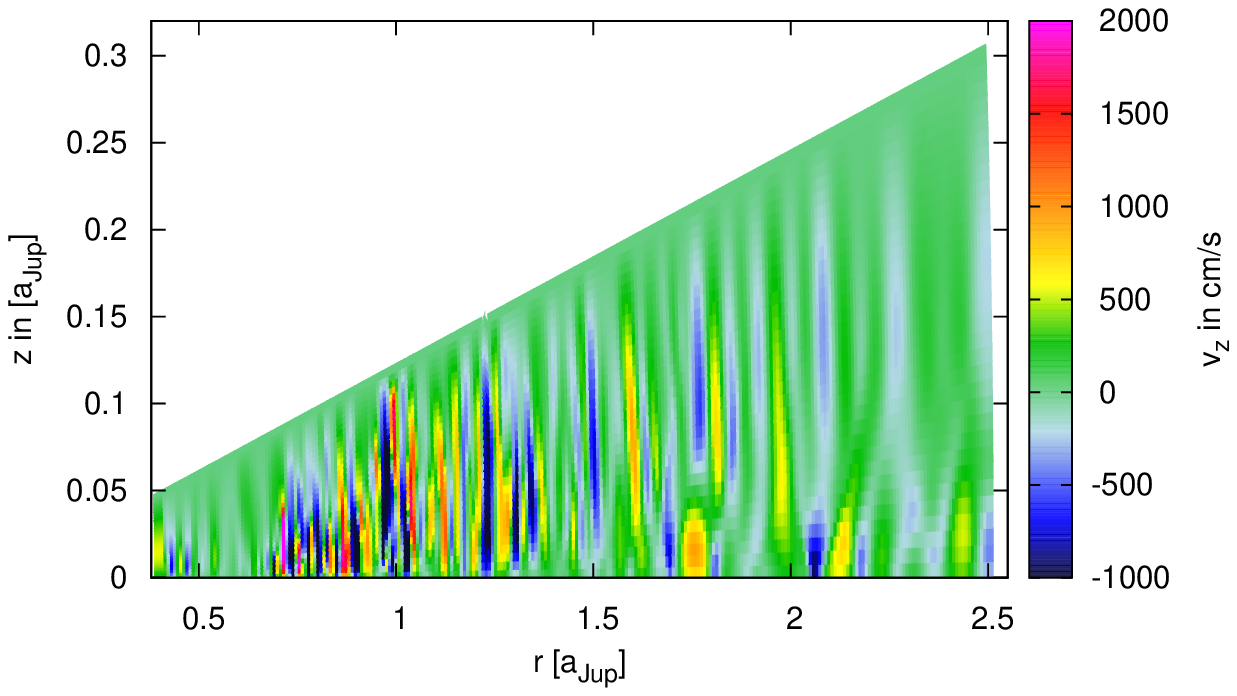}
 \includegraphics[width=0.805\linwx]{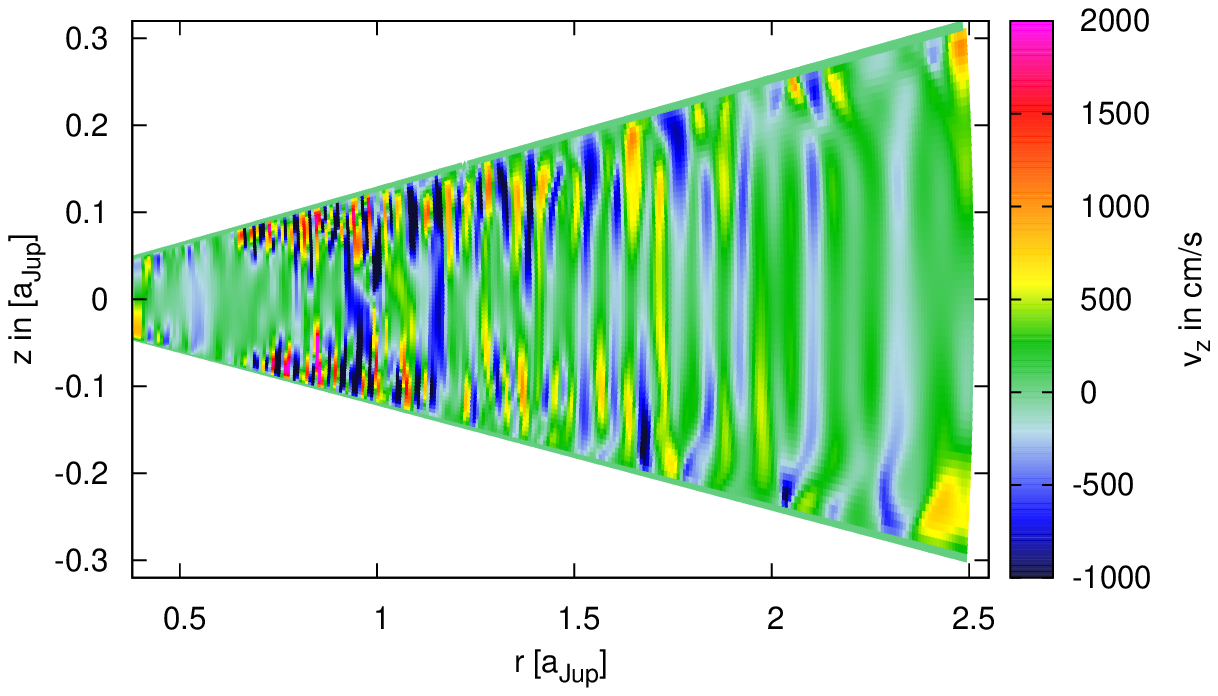}
 \caption{Velocities in $z$-direction for $0.04 M_\odot$ discs with embedded $20 M_{Earth}$ planet. In the top panel only the upper half of the disc was simulated, while in the bottom panel both sides of the disc were simulated (with twice the number of grid cells in $\theta$ direction). The other simulation parameters are identical.
   \label{fig:Mass2DPlanetthetavy}
   }
\end{figure}

When putting a $20 M_{Earth}$ planet in the $0.04 M_\odot$ discs, the structure of the convective cells changes as the disc gets disturbed by the planet. Without the planet, the convective cells are very regular for distances to the central star exceeding $r > 1.25 a_{Jup}$. With an embedded planet these regular structures break down and become very irregular, as can be seen in Fig.~\ref{fig:Mass2DPlanetthetavy}. This effect may caused by to the wake created by the planet inside the disc, which acts as an additional heat source. At shorter distances to the central star, the structure is irregular with or without an embedded planet.

The velocity patterns for the one sided and two sided discs show a little difference. It seems that the fluctuations in velocity are more centred in midplane for the one sided disc, while they seem to be located near the upper and lower boundaries for the two sided discs. This effect has several reasons. In the one sided disc, material is reflected at the midplane of the disc, which might lead to an increase of fluctuations near the midplane. In the two sided disc, material can flow through the midplane, so that the fluctuations near the midplane are reduced. Furthermore, the one sided disc might be unrealistic if convection is present in the disc.

However, the general structure changes when both halves of the disc are computed, independent of an embedded planet. The convection cells are now moving through the midplane of the disc, which was not possible for simulations of only one half of the disc, see also \citet{1993ApJ...416..679K,1999ApJ...514..325K}. Therefore, the surface density structure in the two sided disc (with $M_{disc}=0.04 M_\odot$) is slightly different compared to the one sided disc. In the two sided case, the fluctuations in the surface density continue only to $\approx 1.5 a_{Jup}$, while they covered the whole disc in the one sided case. The temperature profiles, on the other hand, show no difference at all. We therefore only expect little change in the torque acting on planets embedded in one or two sided discs at $r_P=1.0 a_{Jup}$.

Simulations of embedded $20 M_{Earth}$ planets in fully radiative discs that cover both sides of the disc show only very small differences in the velocity pattern in mid-plane of the disc. The fluctuations in time of the torque acting on the planet embedded in $M_{disc} = 0.04 M_\odot$ and $M_{disc} = 0.03 M_\odot$ discs are comparable (with only small differences in the amplitude of the fluctuations) for both simulations, confirming our previous assumptions. Simulations of planets embedded in lower mass discs show no difference at all (simulations not displayed here), because the convective region does not reach to the planet at all.

Considering that a longer distance from the central star resulted in a turn from outward to inward migration for a $20 M_{Earth}$ planet (see Fig.~\ref{fig:MigTorqueMass}) because of a reduction in temperature and density at the given location of the planet, one might argue that the zero-torque distance from the central star might be at longer distances for higher disc masses. Moreover, as the disrupting convective zone in massive discs reaches farther out from the star, outward migration might be possible at larger radii in more massive discs, because the disc's convective zone stops at longer distance to the central star and the density is still high enough to produce the surface density patter needed for outward migration.

\begin{figure}
 \centering
 \includegraphics[width=0.9\linwx]{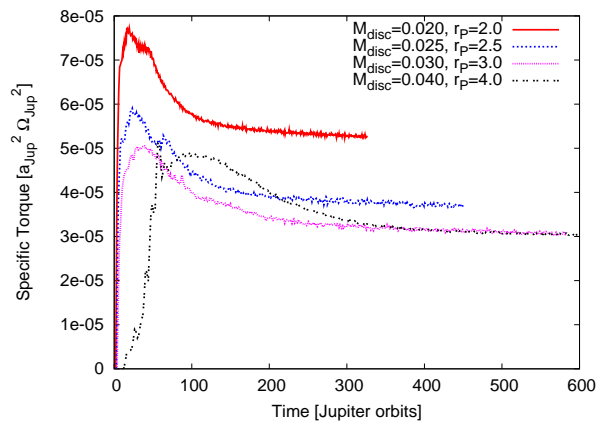}
 \caption{Specific torque acting on $20 M_{Earth}$ planets embedded in $0.02$, $0.025$, $0.03$ and $0.04 M_\odot$ discs at the following distances $r_p=2.0$, $2.5$, $3.0$ and $4.0 a_{Jup}$, respectively.
   \label{fig:MdiscRplanet}
   }
\end{figure}

Additional simulations with $20 M_{Earth}$ planets in $0.02$, $0.025$, $0.03$ and $0.04 M_\odot$ discs with $r_p=2.0$, $2.5$, $3.0$ and $4.0 a_{Jup}$, respectively, confirm our assumption (see Fig.~\ref{fig:MdiscRplanet}). The torques acting on those planets are positive, indicating outward migration, and show no fluctuations in time. The surface density plots also show no sign of convection in the disc at the location of the planet (not displayed here). It seems that outward migration is therefore possible to farther distances from the central star in more massive discs.

The picture of convection in our disc would change when including stellar irradiation because it would heat the surfaces of the disc in contrast to the applied cooling right now. This would result in less convection in the disc. Because the convective region is a result of the higher surface density (increasing $\tau_{rad}$) and viscosity in the disc, a reduction in viscosity could prevent convection in the disc. However, a reduction of viscosity also reduces the torque of an embedded planet, so that outward migration might not be possible any more, even for low-mass discs. The influence of viscosity will also be addressed in a next paper in much more detail.

In self-gravitating discs, the torque acting on an embedded planet can differ by a factor of two compared to non self-gravitating discs, as shown by \citet{2008ApJ...678..483B}. These authors also state that self-gravity has no effect on the corotation torque in the linear regime, but our 3D simulations are in the non-linear regime. Therefore the influence of self-gravity on planet migration in fully radiative discs should be investigated in the future.

The Toomre stability criterion can be used to estimate the stability of discs against self-gravity \citep{1964ApJ...139.1217T}. The stability parameter reads
\begin{equation}
\label{eq:toomre}
   Q = \frac{c_s \kappa_{ep}}{\pi \Sigma G} \ ,
\end{equation}
where $c_s$ is the sound speed in the disc, $\kappa_{ep}$ is the epicyclic frequency, which for Keplerian discs is approximately equal to the
angular frequency $\Omega$, $\Sigma$ is the surface mass density and $G$ is the gravitational constant. In order to achieve stability in discs, 
the stability parameter must be $Q \gg 1$. For all disc masses used in this work, this criterion is fulfilled well, so that the discs are not gravitationally unstable.

Because convection is a 3D effect, 2D simulations (in $r$-$\phi$ direction in the midplane) of fully radiative discs with high discs masses ($M_{disc} > 0.02 M_\odot$) cannot capture this effect. Therefore planets embedded in these 2D simulations will not be exposed to these fluctuations and might therefore be inaccurate near the central star because of convection in the disc.

\section{Summary and conclusions}
\label{sec:Sumcon}

We performed full 3D radiation hydrodynamical simulations of low-mass planets embedded in accretion discs at different distances to the central star and for various disc masses. 

In the first sequence of our simulations we changed the planetary distance to the central star of embedded planets on circular orbits. With increasing distance to the central star, the torque acting on $20 M_{Earth}$ planets embedded in fully radiative discs becomes even more reduced and it reaches negative torques for longer distances. We find an equilibrium, zero-torque distance, to the central star for $20 M_{Earth}$ planets at $r \approx 2.4 a_{Jup}$. This equilibrium distance varies with the planetary mass (for $25 M_{Earth}$ planets it is $r \approx 1.9 a_{Jup}$ and $r \approx 1.4 a_{Jup}$ for $30 M_{Earth}$ planets), indicating that a quite extended region in the disc might act as a feeding zone to create even larger planetary cores. The concept of equilibrium radius (zero torque radius) for planetary embryos in fully unsaturated discs has been stated in \citet{2010ApJ...715:L68} as well and it could easily act as a feeding or collection zone for planetary embryos.

Planets embedded in fully radiative discs migrating outwards create a very sensitive pattern in the surface density distribution. Ahead and inside of the planet a density increase is visible (see second panel from the top in Fig.~\ref{fig:Mig2DRho}), which shrinks with increasing distance to the central star. This density enhancement is indeed accountable for the positive torque acting on the planet (also visible as a spike in the radial torque density distribution in Fig.~\ref{fig:MigGamma3D}), but as the distance to the central star increases, this effect becomes less, so that it cannot overcompensate the negative Lindblad torques any more, which results in inward migration. 

We compared our results to the recently developed torque formulae by \citet{2010MNRAS.401.1950P,2011MNRAS.410..293P} and \citet{2010ApJ...723.1393M}.
\citet{2010MNRAS.401.1950P} includes just the fully unsaturated torques in the inviscid and adiabatic case, shows no torque reversal option for constant $\beta$ and is as such unphysical when comparing it with the long-term evolution of planets. 
However, as in our simulations $\beta$ changes in the disc with distance to the central star ($\beta (r=1.0 a_{Jup}) =1.7$ to $\beta (r=5.0 a_{Jup}) = 0.33$), the torque given by the formula becomes negative for large distances to the central star. But, overall the match of the torque of the formula with our simulations is not good. This formula is only valid in the first orbits after the planet is embedded when the torques are still unsaturated, however, the torques do saturate in time. 

However, as expected, the improved version of \citet{2011MNRAS.410..293P} that includes viscous and heat diffusion describes our results more 
accurately. It also features a transition from positive to negative torques, but the negative torques at large distances to the central star are about a factor of two larger than in our simulations. Even though an exact match of the torques has not been achieved, the trend seems to be caputured quite well by the formula.

The torque formulae are derived for 2D discs with thermal diffusion operating only in the disk's midplane,
while our simulations are 3D, and take full account of vertical diffusion as well, which can change the structure of the disc.
The formulae were also derived for given gradients in temperature and surface density, but in a real disc the temperature and surface density profiles are disturbed when a planet is embedded in a disc. Because the formulae were derived and checked for a $5 M_{Earth}$ planet (with about $20$ per cent agreement), the disturbances of such a small planet in a disc are much weaker than for our embedded $20 M_{Earth}$ planet, which may give rise to more significant, non-linear disturbances in the temperature and density profiles. All of this may lead to differences between our simulations and the theoretical formulae in the torque acting on the planet.

The formula of \citet{2010ApJ...723.1393M} does not match our simulations that well. 
The torques from the formula are always negative, which is in contrast to our simulations, where we find outward migration for $r<2.5 a_{Jup}$.
For long distances to the central star, \citet{2010ApJ...723.1393M} match better than \citet{2011MNRAS.410..293P}, as the Lindblad torque is reduced.
The Lindblad torque of \citet{2010ApJ...723.1393M} also matches quite well with the isothermal torque of \citet{2010ApJ...724..730D},
when taking the factor $\gamma$ into account due to the different sound speeds in isothermal and fully radiative discs.
The overall trend to have torques from positive to negative seems best achieved with \citet{2011MNRAS.410..293P},
but a quantitative agreement is still lacking.
In Appendix \ref{app:comp} we discuss the influence of the smoothing length on the formulae.

For increasing disc masses, the temperature, density (in midplane), and aspect ratio of the disc increases in the equilibrium state where  viscous heating and radiative transport/cooling are in balance. The convective zone in the inner discs stretches farther out with increasing disc mass, resulting in high fluctuations of the surface density in our computed domain for discs with a mass higher than $M_{disc} \approx 0.02 M_\odot$. 

Starting from a $M_{disc}=0.01 M_\odot$ disc, the torque acting on embedded $20 M_{Earth}$ planets decreases for increasing disc masses. As the disc mass increases, the convective zone in the disc stretches farther out from the central star and influences planetary migration. The fluctuations in the disc's density disrupt the torque acting on the planet on a stationary orbit for high-mass discs in a way that the torque is very irregular and shows high fluctuations as well, making it difficult to determine the correct direction of migration. For lower disc masses, the torque reduces as well, presumably because of the same reasons as the torque is reduced for longer distances to the central star in a discs with $M_{dics} = 0.01 M_\odot$.

The formula in \citet{2011MNRAS.410..293P} fits within a factor of three with our 3D simulations for planets in discs with $M_{dics} \approx 0.01 M_\odot$. For higher disc masses, the difference between the formula and our 3D simulations becomes smaller. 
However, the 3D simulations show a decreasing torque for increasing disc mass, while the \citet{2011MNRAS.410..293P} formula increases slightly.
As the disc mass increases, viscous heating increases and cooling becomes inefficient, which results in a structure similar to an adiabatic disc.
In adiabatic discs the corotation torques saturates, resulting in a lower torque acting on the planet, hence the drop of the torque for increasing disc masses.
Convection certainly plays a role in more massive discs, but it is unaccounted for in \citet{2011MNRAS.410..293P}. In more massive discs the convective zone reaches longer distances from the central star, disrupting the density pattern near the embedded planet and thus creating fluctuations in the torque acting on the planet. These disruptions in the density pattern are caused by the convective cells evolving in the disc. These cells also change in time, giving rise to the stronger fluctuations of the torque.

Convection is inefficient for transporting angular momentum \citep{1993ApJ...416..679K,2010MNRAS.404..L64}, but the influences of convection on planetary migration are very dramatic, because a planet close to the star in the convective zone of the disc is essentially disrupted. The direction of migration is not clearly determinable any more, but when the planet is farther out in the massive disc and the convection fades away, the direction of migration is easy to specify, indicating outward migration. Therefore the zero-torque radius for migration lies farther out in more massive discs.

Convection is also a 3D effect only and cannot be simulated in 2D discs (in $r$-$\phi$ direction in the midplane). Two-dimensional simulations of planets in massive discs ($M_{disc} > 0.02 M_\odot$) might therefore be inaccurate near the central star, because the effects of convection are not considered.

\appendix
\section{Comparison with \citet{2011MNRAS.410..293P}}
\label{app:comp}

In Section (\ref{sec:outwardrange}) we compared our numerical results to the analytical torque formula derived in \citet{2011MNRAS.410..293P}.  As the derivation is rather cumbersome, we present here a brief summary of the relevant contributions  to the total torque acting on a planet, so that the reader can follow our calculations. In our notation we closely follow \citet{2011MNRAS.410..293P}.

The total torque acting on a low-mass planet consists of two main contributions the Lindblad torque, $\Gamma_L$, plus the corotation torque, $\Gamma_c$
\begin{equation}
\label{eq:total}
    \Gamma_{tot} = \Gamma_L + \Gamma_c \ . 
\end{equation}
The Lindblad torque is caused by the action of the induced spiral arms and is given as \citep{2008A&A...485..877P}
\begin{equation}
\label{eq:lindblad}
     \gamma \Gamma_L / \Gamma_0 = -2.5 - 1.7 \beta + 0.1 \alpha \ ,
\end{equation}
where $\alpha$ denotes the negative slope of the surface density profile $\Sigma \propto r^{-\alpha}$, $\beta$ refers to the slope of the temperature profile  $T \propto r^{-\beta}$, and $\gamma$ is the adiabatic index of the gas.

It is important to notice that all torques listed here are normalized to 
\[
      \Gamma_0 = \left(\frac{q}{h}\right)^2 \Sigma_P r_p^4 \Omega_P^2 \ ,
\]
with $q$ the planet/star mass ratio, $\Sigma_P$ the surface density at the planet's location, $\Omega_P$ the rotation frequency of the planet in the disc,
and $h$ is in this appendix the {\it isothermal} relative disk height. 

The corotation torque is now split into the barotropic part and an entropy-related part:
\[
    \Gamma_c = \Gamma_{c,baro} + \Gamma_{c,ent} \ ,
\]
where the first part applies to barotropic flows where the pressure only depends on the density, and it depends on the gradient of the vorticity in the flow; the second part relates to the variations of entropy. Each of them is split again into a linear contribution and a so-called horseshoe drag contribution. This separation is necessary because the two parts are affected differently by the diffusion processes. The barotropic part of the (non-linear) horseshoe drag is given by
\begin{equation}
\label{eq:hsbaro}
    \gamma \Gamma_{hs,baro} / \Gamma_0 = 1.1 ( 1.5 - \alpha)
\end{equation}
and the entropy-related part of the horseshoe drag is given by
\begin{equation}
\label{eq:hsent}
    \gamma \Gamma_{hs,ent} / \Gamma_0 = 7.9 \frac{\xi}{\gamma} ,
\end{equation}
where $\xi = \beta - (\gamma - 1.0) \alpha$ is the negative of the power-law index of the entropy. 
We note that the total torque formula given by \citet{2010MNRAS.401.1950P}, as summarized in Eq.~(\ref{eq:paar09}), is exactly the sum
$\Gamma_{tot} = \Gamma_L + \Gamma_{hs,baro} + \Gamma_{hs,ent}$.

The barotropic part of the linear corotation torque reads as
\begin{equation}
\label{eq:clinbaro}
    \gamma \Gamma_{c,lin,baro} / \Gamma_0 = 0.7 (1.5 - \alpha) ,
\end{equation}
and the entropy-related part of the linear corotation torque is given by
\begin{equation}
\label{eq:clinent}
    \gamma \Gamma_{c,lin,ent} / \Gamma_0 = \big(2.2 - \frac{1.4}{\gamma} \big) \xi .
\end{equation}

Owing to the difference between the isothermal and adiabatic sound speed, differences in the torque arise. To compensate for this, the adiabatic index $\gamma$ should be replaced by an {\it effective} $\gamma$:
\begin{eqnarray}
     \gamma_{eff} = \frac{2Q \gamma}{\gamma Q + \frac{1}{2} \sqrt{2 \sqrt{(\gamma^2 Q^2 + 1)^2 - 16 Q^2 (\gamma -1)} + 2 \gamma^2 Q^2 -2}} \nonumber \ ,
\end{eqnarray}
so that all $\gamma$'s in the previous equations (\ref{eq:lindblad} to \ref{eq:clinent}) have to be replaced by $\gamma_{eff}$. The parameter $Q$ is given by
\[
    Q=\frac{2 \chi_P \Omega_P}{3 h c_s^2} = \frac{2 \chi_P}{3 h^3 r_P^2 \Omega_P} \ ,
\]
where $h=H/r$ and 
\[
\chi_P = \frac{16 \gamma (\gamma -1) \sigma_P T_P^4}{3 \kappa_P \rho_P^2 H_P^2 \Omega_O} \ ,
\]
with $\kappa$ beeing the opacity and $\sigma$ the Stefan-Boltzmann constant. Please note that in \citet{2011MNRAS.410..293P} a factor of 4 is missing.
The final correction relates to the non-ideal effects of viscosity and heat transfer, which both have to be present to avoid the saturation of the corotation torque. The barotropic part of the horseshoe drag is not affected by thermal diffusion and is only determined by the viscosity. According to \citet{2011MNRAS.410..293P}  it can be written as
\[
    \Gamma_{c,baro} = \Gamma_{hs,baro} F(p_\nu) G(p_\nu) + (1 - K(p_\nu)) \Gamma_{c,lin,baro} ,
\]
where $\Gamma_{hs,baro}$ and $\Gamma_{c,lin,baro}$ are given by equations (\ref{eq:hsbaro} and \ref{eq:clinbaro}), but now with $\gamma \to \gamma_{eff}$. $F(p)$ (eq.~\ref{eq:Fp}) governs saturation and $G(p)$ (eq.~\ref{eq:Gp}) and $K(p)$ (eq.~\ref{eq:Kp}) govern the cut-off at high viscosity.

For the non-barotropic, entropy-related corotation torque \citet{2011MNRAS.410..293P} find
\begin{eqnarray}
      \Gamma_{c,ent} &= \Gamma_{hs,ent} F(p_\nu) F(p_\chi) \sqrt{G(p_\nu) G(p_\chi)} \newline \nonumber \\
      &+ \sqrt{(1-K(p_\nu))(1-K(p_\chi))} \Gamma_{c,lin,ent} \nonumber \ ,
\end{eqnarray}
where $\Gamma_{hs,ent}$ and $\Gamma_{c,lin,ent}$ are given by equations \ref{eq:hsent} and \ref{eq:clinent}, again with $\gamma \to \gamma_{eff}$, and $p_\chi$ is the saturation parameter associated with thermal diffusion.

The function $F(p)$ is given by
\begin{equation}
\label{eq:Fp}
    F(p) = \frac{1}{1+ (p/1.3)^2} \ .
\end{equation}

The function $G(p)$ is given by
\begin{equation}
\label{eq:Gp}
G(p) =  \left\{
    \begin{array}{cc} 
    \frac{16}{25} \big( \frac{45 \pi}{8} \big)^{3/4} p^{3/2}
    \quad &  \mbox{for} \quad p < \sqrt{\frac{8}{45 \pi}}  \\
    1 - \frac{9}{25} \big( \frac{8}{45 \pi} \big)^{4/3} p^{-8/3}
    \quad & \mbox{for} \quad  p \geq \sqrt{\frac{8}{45 \pi}} 
    \end{array}
    \right. .
\end{equation}

The function $K(p)$ is given by
\begin{equation}
\label{eq:Kp}
K(p) =  \left\{
    \begin{array}{cc} 
    \frac{16}{25} \big( \frac{45 \pi}{28} \big)^{3/4} p^{3/2}
    \quad &  \mbox{for} \quad p < \sqrt{\frac{28}{45 \pi}}  \\
    1 - \frac{9}{25} \big( \frac{28}{45 \pi} \big)^{4/3} p^{-8/3}
    \quad & \mbox{for} \quad  p \geq \sqrt{\frac{28}{45 \pi}} 
    \end{array}
    \right. .
\end{equation}

The parameters $p_\nu$ and $p_\chi$, relate to the strength of viscosity and thermal diffusivity, and are given by
\begin{eqnarray}
      p_\nu &=  \frac{2}{3} \, \sqrt{\frac{r_P^2 \Omega_P x_s^3}{2 \pi \nu_p}} \nonumber \ , \\
      p_\chi &= \sqrt{\frac{r_P^2 \Omega_P x_s^3}{2 \pi \chi_p}} \nonumber \ ,
\end{eqnarray}
where $\nu_p$ is the kinematic viscosity and $\chi_p$ the thermal conductivity at the planet location.
$x_s$ is the half width of the horseshoe, given by
\[
	x_s = \frac{1.1}{\gamma_{eff}^{1/4}} \big( \frac{0.4}{\epsilon / h} \big)^{1/4} \sqrt{\frac{q}{h}} \ ,
\]
The scaling with $\epsilon / h$ breaks down for small softening ($\epsilon / h < 0.3$).
All these contributions have to be substituted into equation (\ref{eq:total}) to calculate the total torque acting on the planet. In deriving these formulae one has to use a description for the smoothing of the gravitational potential. Here, a standard $\epsilon$ potential has been assumed, and a smoothing length of $\epsilon/h=0.4$ has been made. 

To compare with our simulations, we have used the following parameter in the formulae above
$\alpha = 0.5, \beta = 1.7, \xi= 1.5$. To evaluate the viscosity parameter $p_\nu$ we use $\nu = 10^{-5}$.
Please note that $\nu_p \approx \chi_p$ for discs in radiative equilibrium.

In \citet{2009A&A...506..971K} we pointed out that a different planetary smoothing results in different torques. This phenomenon can be up to a factor of two for the $\epsilon$ potential with $r_{sm}=0.5$ and the cubic potential with $r_{sm}=0.5$. In Section (\ref{sec:outwardrange}) we compared our numerical simulations to the smoothing with  $\epsilon / h = 0.4$.

\begin{figure}
 \centering
 \includegraphics[width=0.9\linwx]{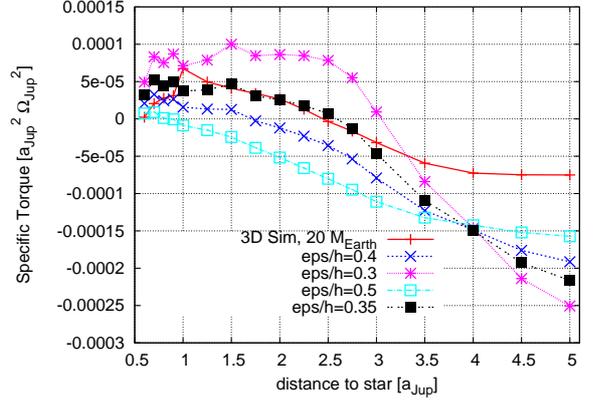}
 \caption{Specific torque acting on $20 M_{Earth}$ planets embedded in a $0.01 M_\odot$ disc. Overplotted are the results of \citet{2011MNRAS.410..293P}
 using various smoothings lengths from $\epsilon / h = 0.3$ and $\epsilon / h = 0.5$.
   \label{fig:MigTorquebh036}
   }
\end{figure}

Because our cubic potential with $r_{sm}$ is deeper in the vicinity of the planet, we used $\epsilon / h = 0.3$, $\epsilon / h = 0.35$ and $\epsilon / h = 0.5$ for the \citet{2011MNRAS.410..293P} formula as well. The results are shown in Fig.~\ref{fig:MigTorquebh036}.
In the $\epsilon / h = 0.5$ case, the torque is negative for all planetary distances, which is not a match at all for our simulations,
but interestingly agrees better with \citet{2010ApJ...723.1393M}.
For $\epsilon / h = 0.3$, the torque is much higher in the positive torque regime and much lower in the negative torque regime. For the $\epsilon / h = 0.35$ case, the formula matches our simulations quite well in the positive torque regime ($r<2.5 a_{Jup}$).
However, for larger distances to the central star, the torque of the formula is about a factor of two to four larger than the torque from our simulations.
Nevertheless, the match for $r<2.5 a_{Jup}$ is quite good.
Obviously, the smoothing of the planetary potential seems to have a huge effect on the torque acting on the planet.
The effect seems much stronger in the formula compared to different smoothings in our 3D hydrodynamical simulations \citep{2009A&A...506..971K} but
they are based on 2D simulations, where a larger impact is to be expected.

\begin{acknowledgements}

B. Bitsch has been sponsored through the German D-grid initiative. W. Kley acknowledges the support through the German Research Foundation (DFG) through grant KL 650/11 within the Collaborative Research Group FOR 759: {\it The formation of Planets: The Critical First Growth Phase}. The calculations were performed on systems of the Computer centre of the University of T\"ubingen (ZDV) and systems  operated by the ZDV on behalf of bwGRiD, the grid of the Baden  W\"urttemberg state. Very helpful discussions with K.M. Dittkrist about the torque formulae are thankfully acknowledged. Finally, we gratefully acknowledge the helpful and constructive comments of an anonymous referee.

\end{acknowledgements}

\bibliographystyle{aa}
\bibliography{kley8}
\end{document}